\documentclass[preprint,superscriptaddress,nofootinbib,preprintnumbers,amsmath,amssymb]{revtex4-1}

\usepackage{amsfonts}
\usepackage{mathrsfs}
\usepackage{leftidx}
\usepackage{amssymb}
\usepackage{placeins}
\usepackage{relsize}
\usepackage{slashed}
\usepackage{multirow}
\usepackage{array, makecell}
\setcellgapes{7.5pt}
\usepackage{xfrac}
\usepackage[dvipsnames]{xcolor}

\usepackage[colorlinks]{hyperref}
\usepackage{epsfig}
\usepackage{graphicx}               
\usepackage{url}
\usepackage{float}

\newcommand{\be}{\begin{equation}}
\newcommand{\ee}{\end{equation}}
\newcommand{\bea}{\begin{eqnarray}}
\newcommand{\eea}{\end{eqnarray}}
\newcommand{\beq}{\begin{eqnarray}}
\newcommand{\eeq}{\end{eqnarray}}
\def\bit{\begin{itemize}}
\def\eit{\end{itemize}}
\def\ben{\begin{enumerate}}
\def\een{\end{enumerate}}
\newcommand{\Fig}[1]{Fig.~\ref{#1}}
\newcommand{\Eq}[1]{Eq.~(\ref{#1})}

\newcommand{\Sec}[1]{Sec.~\ref{#1}}

\newcommand{\App}[1]{App.~\ref{#1}}

\def\mxbar{{\bar{m}_X}}
\def\syn{{\rm syn}}

\newcommand\DN[1][\relax]{%
\ifx\relax#1\relax\else{}^{#1}\fi \!X}

\newcommand{\BE}{\textrm{BE}}
\newcommand{\DM}{\textrm{DM}}
\newcommand{\GeV}{{\rm GeV}}
\newcommand{\cm}{{\rm cm}}
\newcommand{\gram}{{\rm g}}
\newcommand{\sat}{{\rm sat}}

\newcommand{\kfo}{N_{\rm fo}}
\newcommand{\kfobn}{N_{\rm fo}^{*}}

\newcommand{\sphi}{\varphi}
\newcommand{\sV}{\upsilon}
\newcommand{\SIDM}{\sfrac{\sigma_\DM}{m_\DM}}

\begin{document}


\title{Astrophysical Signatures of Asymmetric Dark Matter Bound States}
\author{Moira I. Gresham}
\affiliation{Whitman College, Walla Walla, WA 99362}
\author{Hou Keong Lou}
\affiliation{Theoretical Physics Group, Lawrence Berkeley National Laboratory, Berkeley, CA 94720}
\affiliation{Berkeley Center for Theoretical Physics, University of California, Berkeley, CA 94720}
\author{Kathryn M. Zurek}
\affiliation{Theoretical Physics Group, Lawrence Berkeley National Laboratory, Berkeley, CA 94720}
\affiliation{Berkeley Center for Theoretical Physics, University of California, Berkeley, CA 94720}
\affiliation{Theoretical Physics Department, CERN, Geneva, Switzerland}

\begin{abstract} 
Nuggets---very large stable bound objects arising in the presence of a sufficiently attractive and long-range force and in the absence of a dark Coulomb force---are a smoking gun signature for Asymmetric Dark Matter (ADM).  The cosmology of ADM nuggets is both generic and unique: nuggets feature highly exothermic fusion processes, which can impact the shape of the core in galaxies, as well as give rise to rare dark star formation.   We find, considering the properties of nuggets in a generic extended nuclear model with both attractive and repulsive forces, that self-interaction constraints place an upper bound on nugget masses at the freeze-out of synthesis in the ballpark of $M_{\rm fo} \lesssim 10^{16}$ GeV.  We also show that indirect detection strongly constrains models where the scalar mediator binding the nuggets mixes with the Higgs.  
\end{abstract}

\maketitle
\tableofcontents


\section{Introduction}


Asymmetric dark matter (ADM) \cite{Kaplan:2009ag,Zurek:2013wia,Petraki:2013wwa} with an attractive force may give rise to bound states, called nuggets \cite{Wise:2014ola,Wise:2014jva}.  Depending on the relative strength of the attractive and repulsive forces binding the nugget, nuggets can grow to be quite large, with millions or more constituents per bound state \cite{Hardy:2014mqa} (see also \cite{Krnjaic:2014xza}).  Such large bound states could give rise to new direct detection signatures \cite{Hardy:2015boa}, requiring novel direct detection techniques. In addition, complementarity between direct, indirect, structure formation, and collider constraints can differ substantially from the standard WIMP paradigm, leading to new Dark Matter (DM) model building possibilities. Examples of the striking implications for DM phenomenology when a substantial component is in the form of bound states can be found in mirror dark matter \cite{Khlopov:1989fj,Berezhiani:1995am,Mohapatra:2000qx,Mohapatra:2001sx,Foot:2004pa}, WIMPonium \cite{Pospelov:2008jd,MarchRussell:2008tu,Shepherd:2009sa}, atomic dark matter \cite{Kaplan:2009de,CyrRacine:2012fz}, dissipative dark matter \cite{Fan:2013tia,Fan:2013yva}, and dark nuclei \cite{Detmold:2014qqa} scenarios.

Large bound states generally require fermionic constituents to provide a stabilizing pressure, and must be composed of an asymmetric component for synthesis to be efficient in the early Universe. Thus, large dark matter bound states are a smoking gun signature of fermionic asymmetric dark matter.  Conversely, the existence of large bound states is generic within an ADM scenario; in particular a light mediator of strong self-interactions serves as an effective annihilation channel for depleting the symmetric DM component in the early Universe \cite{Kaplan:2009ag,Lin:2011gj}. If that mediator is a scalar, the self-interactions are attractive and lead to nuggets. 

One may wonder why such large bound states do not arise in the Standard Model (SM).  In the SM, attractive nuclear forces are effectively mediated by pseudoscalar and  scalar bound states, such as the pion and the $\sigma$.  There, however, arbitrarily large nuclei are not synthesized because of the presence of bottlenecks in the early Universe, and, more importantly, because of the presence of a Coulomb barrier.  As we argue quantitatively in \App{app:sm_bottleneck}, the absence of a Coulomb barrier makes an enormous difference in the predicted size of synthesized bound states by permitting fusion at small velocities.  Furthermore, analogs of the strong $A=8$ bottleneck in the absence of a Coulomb force may easily be circumvented in a more general nugget model, as $^8$Be is only barely unstable. We conclude that with very modest modifications to the structure of the hidden sector relative to the SM, the synthesis of very large composite states of ADM could proceed unblocked, though this will require solving low-$N$ bound state problems to verify.

Large bound states, characterized typically by $N > 10^4$ constituents, are interesting to consider as a DM candidate because their observational signatures---from early Universe cosmology, to impacts on the formation of DM halos, to direct and indirect detection---are quite distinct from other DM candidates that have been widely studied:  
\begin{itemize}
\item As we argue in \Sec{subsec:synthesis}, once one proceeds past the low-$N$ dark nuclei, the size of the synthesized nuggets is quite insensitive to the UV physics of the model, and instead depends on only a few infrared parameters.  This fact allows us, in combination with astrophysical constraints, to make general statements about the size of the nuggets, targeting the features of the nuggets relevant for searches.
\item Nuggets, being large composite states, tend to have large self-interactions, which can impact the shapes of DM halos in the late Universe.
Unlike the standard Self-Interacting DM (SIDM) scenario, however, nuggets are generally as likely to interact by fusing as they are to elastically scatter.
Fusion is of course highly inelastic and, in the class of models we consider, remains exothermic up to arbitrarily large size; cold fusion is realized in these models due to the absence of the analog of electromagnetism. Nugget self-interactions can lead to accelerated mass aggregation at galactic centers, which may provide an efficient way to feed supermassive blackholes.
\item The exothermic and dissipative fusion reactions allow for the possibility of star formation in early protohalos.
\item The byproducts of a single fusion interaction can include (many) force mediators and/or nugget fragments analogous to the common byproducts of SM nuclear interactions: photons, alpha particles, and neutrons. If the fusion byproducts are allowed to decay to SM final states, the observed flux of photons in the galaxy may place a constraint on these models. 
\item Nuggets are extended, massive objects, whose direct detection signals are different from those of WIMPs.
\end{itemize}

In \cite{Gresham:2017zqi, Gresham:2017cvl}, we explored the properties and synthesis of nuggets, focusing on the most deeply bound nuggets with only a scalar mediator. Models with DM coupled only through a light scalar mediator contain the minimal matter content necessary to assemble large ADM bound states; the light mediator is solely responsible for binding both the large and small nuggets, and for allowing the first step of synthesis to proceed kinematically---the analog of deuterium formation, which proceeds through photon emission. Thus this minimal model is fairly predictive but also restrictive. 

Here we consider a more general scenario. 
In general, the size nuggets at freeze-out (fo) of synthesis in the early Universe, $N_\text{fo} (M_\text{fo})$, is largely determined by three dimensionful parameters: number density of bound nucleons, $n_\text{sat}$; mass per constituent, $\bar m_X$, of large nuggets; and the nugget synthesis temperature $T_\syn$ \cite{Hardy:2014mqa}. 
 Although many of our results do not depend on details of the model, in order to be explicit, we consider a concrete effective model in which the dark sector contains a conserved and stable fermionic species in addition to multiple species of mediators: vector, pseudoscalar, and pseudovector in addition to a scalar.   
 The addition of such mediator states opens up $(n_\sat, \mxbar, T_\syn)$ parameter space significantly as compared to the scalar-only model studied in \cite{Wise:2014ola,Wise:2014jva,Gresham:2017cvl}. 
For instance, as occurs for nuclear matter, a repulsive vector and attractive scalar interaction can almost cancel one another, leading to a large hierarchy between  binding energy per constituent and the constituent mass. Additionally, spin-dependent pseudoscalar-mediated interactions can decouple properties of small and large bound states, changing $T_\syn$ relative to $\mxbar$.
 
\begin{figure}
{\centering
\includegraphics[width=0.55 \textwidth]{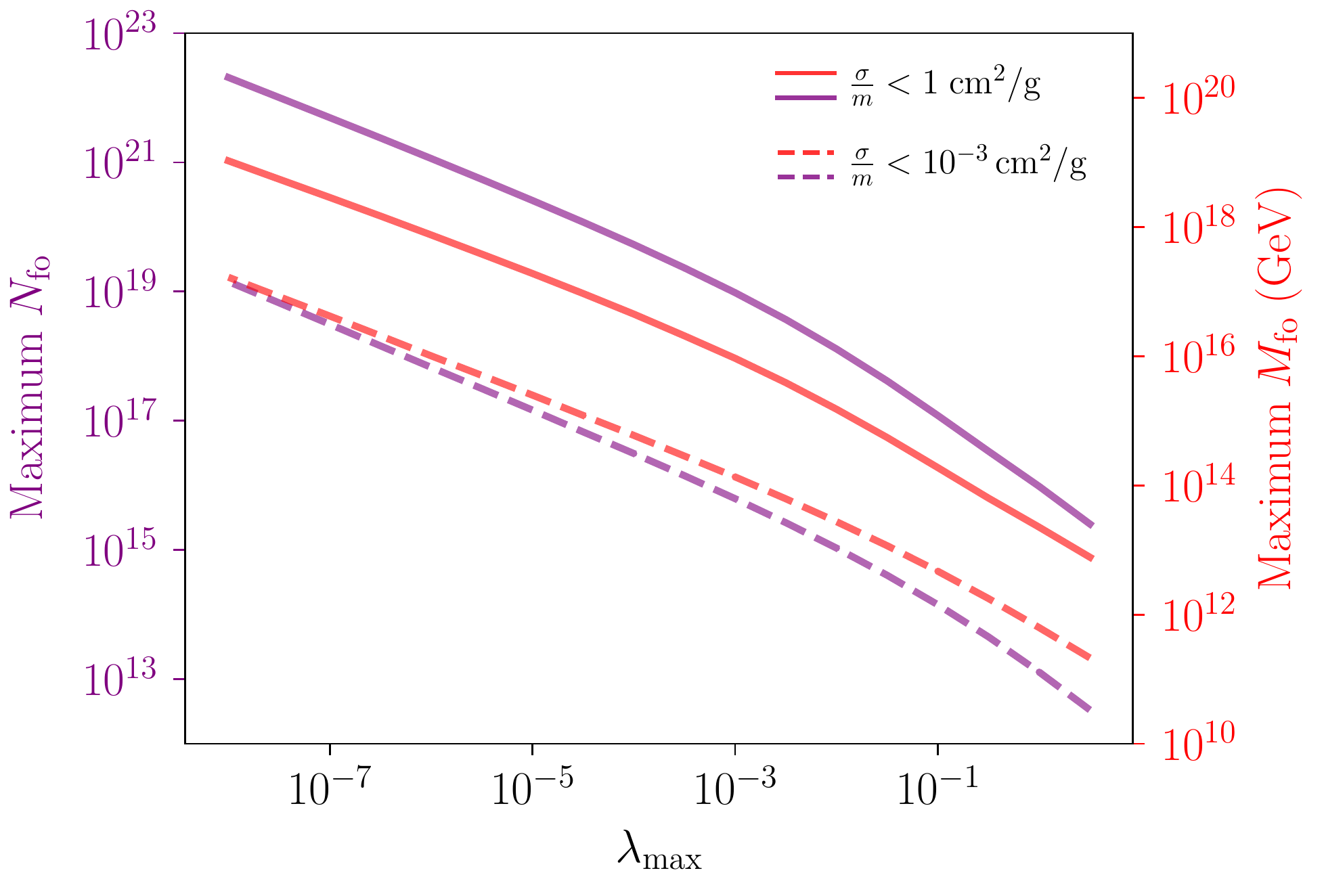}
} 
\caption{Maximum achievable nugget size $\kfo$ (purple) and nugget mass $M_{\rm fo}$ (red) as a function of the maximum scalar force mediator quartic coupling $\lambda_{\max}$. Given a conservative constraint on late-universe DM nugget interaction cross sections $\SIDM \lesssim 1 \; \cm^2/\gram$  (solid) or  $\SIDM \lesssim 10^{-3} \; \cm^2/\gram$ (dashed), maximum possible sizes are realized when only a scalar mediator contributes to large nugget properties (i.e.~when effects of a vector mediator are negligible). $\lambda_{\max}$ serves as a measure of fine tuning for achieving large nuggets, as radiative corrections tend to drive $\lambda$ to be relatively large.  The synthesis temperature is (conservatively) taken to be $T_\syn = (m_X - \mxbar)/15$---the maximum possible 2-body binding energy times a typical Boltzmann suppression factor, $1/30$. A typical model is expected to have lower synthesis temperature and therefore smaller freeze-out sizes and masses.}
\label{fig:Nmax_Mmax}
\end{figure}

One important result of our analysis is generic bounds on the largest possible sizes of DM bound states; these bounds will impact search techniques for nuggets. 
Assuming nuggets are the dominant form of DM, the combination of conservative astrophysical limits on self-interactions discussed in \Sec{sec:interaction}, with general considerations for the nugget properties discussed in \Sec{sec:extended_nuclear}, translate into upper bounds on synthesized nugget size.  
We will show explicitly that large synthesized nuggets require a relatively flat potential for the scalar mediator binding the nugget together, such that a large scalar mean-field can be sustained in a nugget. These constraints are summarized in \Fig{fig:Nmax_Mmax}, as a function of the scalar potential quartic, $\lambda$. Note that the quartic is normalized such that the interaction is given by ${g_\phi^4 \over 3 \pi^2} \frac{\lambda \phi^4}{2}$ (see \Sec{sec:extended_nuclear} for details). Given that there is no symmetry forbidding a quartic term in the potential, models with large synthesized nuggets are tuned. Taking $\lambda \gtrsim 10^{-3}$, a bound $M_{\rm fo} \gtrsim 10^{16}\; \GeV$ is obtained.
 
This paper systematically explores the dominant astrophysical features of, and constraints on, ADM nuggets. In \Sec{sec:extended_nuclear} we will summarize and extend results from previous work on nugget properties and synthesis  that set the foundation for our quantitative analysis of the cosmology and astrophysics of nuggets. 
 Then, in \Sec{sec:interaction} we derive general constraints in $(n_\sat, \mxbar, T_\text{syn})$ parameter space from DM self-interactions, and discuss scenarios where gravothermal collapse of galactic halos are a relevant constraint. 
In Sec. \ref{sec:otherconstraints} we will argue that diffuse X-ray and gamma-ray flux observations constrain models in which mediators can decay to SM particles; this includes the Higgs portal model of \cite{Wise:2014jva}.   We also show that ADM star formation in early protohalos is possible, though rare, in viable regions of parameter space.  Lastly, we discuss how nugget synthesis changes in the presence of a bottleneck similar to the $^8$Be bottleneck in the SM. See Figs.~\ref{fig: saturated nugget pic} and \ref{fig: fusion picture} for summary graphics.


\section{Extended Model for Large Bound States of Fermions}
\label{sec:extended_nuclear}

For very large bound states to be realized in our Universe, (a) large bound state solutions must exist and (b) the bound states must be synthesized efficiently in the early Universe.  
To satisfy (a), we consider fermionic constituents with a generic Lagrangian given by
\begin{align}
  \mathcal{L} &= \bar X(i\slashed\partial - m_X)X +
  \frac{1}{2}
  (\partial \phi)^2 +  \frac{1}{2}
  (\partial a)^2 - \frac{1}{4}V^2_{\mu\nu} - \frac{1}{4}A^2_{\mu\nu} - \frac{1}{2} m_\phi^2\phi^2
-\frac{1}{2} m_a^2 a^2 + \frac{1}{2} m_V^2V_\mu^2
  \notag \\
& \quad - \bar X \left[ (g_\phi \phi + ig_a a \gamma^5) + (g_V \slashed V + g_A \gamma^5 \slashed A) \right]X
- V(\phi, a, V, A) \,.
\label{eq:lagrangian}
\end{align}
In addition to describing bound states of elementary fermions with both vector and scalar force mediators, such a Lagrangian can arise from QCD-like interactions. The scalar $\phi$ and pseudoscalar $a$ are analogous to the isospin singlet $f_0(500)$ (formerly $\sigma$) and $\eta$ mesons, and the vector $V_\mu$ and pseudovector $A_\mu$ are analogous to the isospin singlet $\omega$ and $f_1$  mesons. We ignore a tensor field (the analog of $f_2$) and other higher spin states for simplicity. 
In general, there may be additional flavor indices for all the fields. 
In our regime of interest, where the constituent number is large, relativistic mean field theory (RMFT) is a good approximation and flavor non-singlet fields are expected to have zero expectation values and thus be negligible.\footnote{Large stable SM nuclei violate isospin due to electromagnetism. Absent an analog of electromagnetism, large bound states are flavor symmetric and only flavor singlet fields are important. The effect of flavor is then simply an increase of the fermionic degrees of freedom from $2$ to $2 f$, with $f$ the size of the flavor group.}  
Additionally, we expect the total effect of spin-dependent interactions within very large nuggets to be highly subdominant to that of spin-independent ones, leading to very small expectation values of pseudoscalar and pseudovector fields relative to those of a scalar or vector fields.\footnote{We expect the ground states to be close to spherically symmetric and parity even. For spherically symmetric, parity-even states, $\langle a \rangle$, $\langle A^\mu \rangle$, and  $\langle V^i \rangle$ must all vanish.}
Therefore the pseudoscalar and pseudovector $a$ and $A_\mu$ can be safely ignored in the limit where RMFT applies. This limit of \Eq{eq:lagrangian} is known as the $\sigma-\omega$ model \cite{Walecka:1974qa}, and it well describes the bulk properties (radius and energy density) of large SM nuclei. 

Although $a$ and $A_\mu$ (and flavor non-singlet fields including the analog of pions) are ignored in our large bound state calculations, they could be very important in determining the properties of small-$N$ states. A light $a$ would lead to a more strongly bound $\DN[2]$ state while a light $A_\mu$ could destabilize it. Additionally, $\DN[3]$ or $\DN[4]$ can be destabilized if their binding energies become too small compared to $\DN[2]$, which may lead to strong bottlenecks. Given strong model dependence for small $N$ nugget properties, we will remain agnostic about the dynamics of dark nucleosynthesis for these states, and assume, in the absence of the bottleneck, that synthesis is able to quickly proceed well beyond the size where RMFT calculations are valid. The discussion for synthesis in the presence of a strong bottleneck is reserved for \Sec{sec:bottleneck}, where we assume a small fraction of nuggets are able to squeeze through a strong bottleneck beyond $\DN[2]$. In either case, the small-$N$ physics is roughly parameterized by $T_{\syn}$, the temperature when synthesis begins. 

\subsection{Saturation Properties}

\begin{figure}
\includegraphics[width=0.7\textwidth]{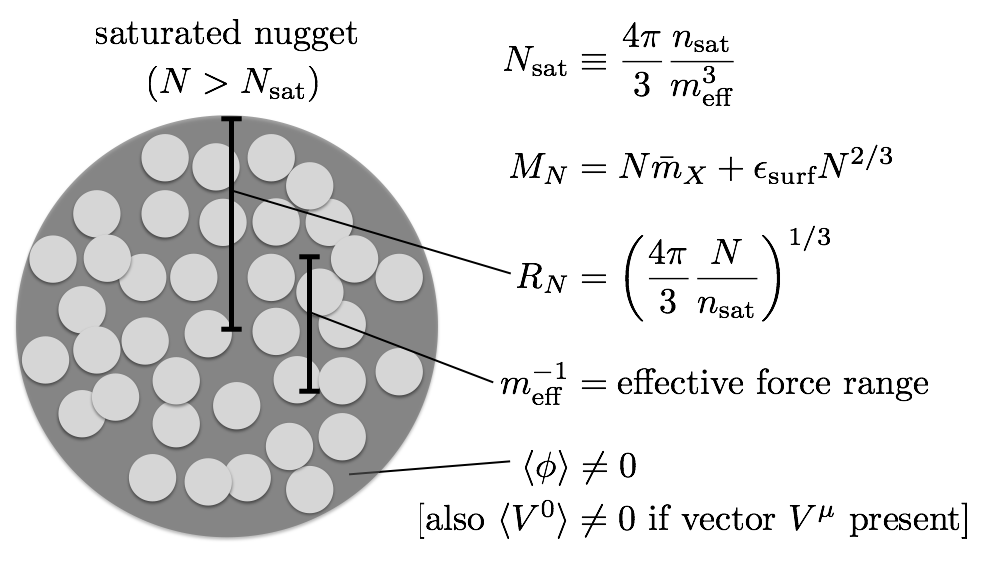}
\caption{Pictorial representation of saturated nuggets: bound states $\DN[N]$ of $N$ fermions, $X$, with $N > N_\text{sat}$. The rest energy per constituent, $\bar m_X$, and density, $n_\text{sat}$, are determined as functions of Lagrangian parameters. See \Eq{eq: CV Cphi definitions} and \Eq{eq: nugget params}. Even if pseudoscalars ($a$) and pseudovectors ($A^\mu$) mediate DM interactions, we expect only a scalar ($\phi$) and vector ($V^\mu$) mediator to contribute to large-$N$ properties. Saturated bound state solutions are generic as long as the scalar interaction is sufficiently strong so that $\bar m_X < m_X$. } \label{fig: saturated nugget pic}
\end{figure}

For a simple scalar-mediator-only model, and using RMFT, we showed in Ref.~\cite{Gresham:2017zqi} that large bound states eventually saturate: their density approaches a constant, $n_\sat$, independent of size, $N$. In this limit, the geometric cross section of a nugget simply scales as
\begin{align}
\sigma_N \sim \pi R_N^2 \simeq \pi \left({4 \pi n_\text{sat} \over 3 }\right)^{-\frac{2}{3}}N^{\frac{2}{3}}\,.
\label{eq: cxn}
\end{align}
As we will justify in \Sec{subsec:cxn}, $\sigma_N$ is also the interaction cross section up to $\mathcal{O}(1)$ factors. We also showed that the saturation limit is valid as long as the nugget size exceeds the force range of the mediator inside the nugget, and that the nugget mass is well described by the liquid drop model,
\begin{align}
M_N = N m_X - \BE_N \approx N \bar m_X + \epsilon_\text{surf} N^{2/3} \,,
\label{eq: liquid drop} 
\end{align}
where $\BE_N$ is the $\DN[N]$ binding energy, $\bar m_X$ is the energy per constituent (the chemical potential) in the $N\rightarrow \infty$ limit, and $\epsilon_{\rm surf}>0$ characterizes the surface energy of the nugget. Total energy per constituent decreases as $\bar m _X + \epsilon_\text{surf} N^{-1/3} $ and therefore it is energetically favorable to form ever larger nuggets. This is in stark contrast to SM nuclei, where nuclei are destabilized (in the sense that fission is exothermic) beyond $^{56}{\rm Fe}$ due to electroweak interactions. In the absence of electromagnetism in the dark sector, we expect saturation properties to hold as long as the nugget size exceeds the effective force range of all mediators inside the nugget---that is, when $N \gtrsim N_\text{sat}$, where
\begin{equation}
N_\text{sat} \equiv {4 \pi \over 3 }{n_\text{sat} \over m_\text{eff}^3}\,.
\end{equation}
For scalar only models, $m_\text{eff} = \sqrt{m_\phi^2+2V(\langle \phi\rangle)/\langle \phi \rangle^2}$ is the effective mass of the scalar inside the nugget. See \Fig{fig: saturated nugget pic} for a summary of saturated nugget parameters.

Since the surface energy $\epsilon_{\rm surf}$ is only relevant when considering details of fusion processes, nugget bound states are well characterized by just two dimensionless quantities $\mxbar/m_X$ and $n_\sat/\mxbar^3$, along with $m_X$ that sets the scale of the system.  In the RMFT approximation, the constituents inside a nugget are described as a free Fermi gas with Fermi momentum $k_F$, with a Dirac mass shifted by the scalar mean-field, $m_* = m_X - g_\phi \langle \phi \rangle$, and a chemical potential ($\bar m_X$) shifted by the vector mean-field. The calculations are detailed in Appendix \ref{app:saturation}, and here we summarize key results (also see {\em e.g.}~\cite{Walecka:1995mi,1996cost.book.....G}). We have
\begin{align}
   \mxbar = g_V \langle V^0 \rangle + \sqrt{k_F^2 + m_*^2}\,, \label{eq: nugget scales}
\qquad 
{\rm and}\qquad
n_{\sat} = \langle X^\dagger X \rangle = g_\text{dof} \int_0^{k_F} {d^3 \vec k \over (2 \pi)^3} = g_\text{dof} \frac{k_F^3}{6\pi^2 }\,,
\end{align}
where $\mxbar$ is the mass per constituent (chemical potential) and $g_\text{dof}=2$ the fermionic degrees of freedom. Binding requires $\bar m_X < m_X$, and thus the effective mass must always be smaller than $m_X$. The vector field equation of motion leads to $ \langle V^0\rangle = {g_V \over m_V^2} \langle X^\dagger X \rangle$, while the scalar field equation of motion relates $k_F$ to $m_*$. Together with the equilibrium condition of zero pressure, $\mxbar$ and $n_\sat$ are determined as functions of Lagrangian parameters.  The saturation density is constrained by the inequality ${n_{\sat}/\mxbar^3 \le g_\text{dof}/(6\pi^2)}$, with the upper bound realizable only in the scalar-only and ultra-relativistic $k_F /m_* \rightarrow 0$ limit.

In Appendix~\ref{app:saturation} we also derive analytic formulas for the nugget properties applicable in the ultrarelativistic limit ($k_F / m_* \gg 1$), which includes regions of large geometric cross section with $n_\sat/m_X^3 \ll g_{\rm dof} / (6 \pi^2)$. They depend on two dimensionless quantities,
\begin{align}
C_V^2 \equiv {g_V^2 \over 3 \pi^2}{m_X^2 \over m_V^2} \qquad \text{and} \qquad C_{\phi}^{2} \equiv  {g_\phi^2 \over 3 \pi^2 }\frac{m_X^2}{m_\phi^2}\left[ 1 + \frac{2 g_\phi^2 V(m_X/g_\phi)}{m_\phi^2 m_X^2} \right]^{-1}\,, \label{eq: CV Cphi definitions}
\end{align}
where we have set $g_\text{dof}=2$ for simplicity, and $V(\phi)$ is the potential for the scalar mediator. We derived analytic formulae for the dimensionless variables $n_\sat/\mxbar^3$ and $\mxbar/m_X$, valid in the regime $C_V^4 C_{\phi}^{-2} \ll 1$ and $C_V^4 C_{\phi}^{-2}\gg 1$, and accurate within 33\% throughout intermediate values: 
\begin{align}
  \frac{n_{\rm sat}}{\mxbar^3} =
                \begin{cases}
                  \frac{1}{3\pi^2} & \frac{C_V^4}{C_{\phi}^{2}} \le \frac{1}{8} \\
                  \frac{1}{3\pi^2}\left[\frac{1}{2}+ \left(\frac{C_V^4}{C_{\phi}^2}\right)^{\frac{1}{3}}\right]^{-3} & \frac{C_V^4}{C_{\phi}^{2}} > \frac{1}{8}
                \end{cases}
\qquad \quad
\frac{\mxbar}{m_X}=
                    \begin{cases}
                      \left(\frac{2}{C_{\phi}^2} \right)^{\frac{1}{4}}
                      & \frac{C_V^4}{C_{\phi}^{2}} \le \frac{1}{8} \\
                       \frac{1}{(C_\phi C_V)^{\frac{1}{3}}}\left[\frac{1}{2}+ \left(\frac{C_V^4}{C_{\phi}^2}\right)^{\frac{1}{3}}\right] & \frac{C_V^4}{C_{\phi}^{2}} > \frac{1}{8}                 
                    \end{cases} \label{eq: nugget params}
\end{align}
As expected from \Eq{eq: nugget scales}, we see that the inclusion of a vector generally decreases $n_{\sat}/\mxbar^3$. In order for the solution to be self-consistent, it must be binding ($\mxbar < m_X$). This is possible as long as $C_V < C_{\phi}$. The approximations break down as $\bar m_X / m_X \rightarrow 1$.

It is instructive to fix a benchmark potential to see explicitly how the nugget parameters $(n_{\sat}/\mxbar^3, \mxbar/m_X)$ are constrained based on Lagrangian parameters. Assuming the scalar potential contains only a quartic term, $V(\phi) = {g_\phi^4 \over 3 \pi^2} \frac{\lambda \phi^4}{2}$, we have $C_{\phi}^{-2} = 3\pi^2 m_\phi^2/(g_\phi^2 m_X^2) + \lambda$. Given that there is no symmetry forbidding the existence of a quartic term, a small $\lambda$ generally requires tuning. Even in the limit where there is an approximate shift symmetry controlled by $g_\phi$, $\lambda$ is expected to be sizable given our choice of normalization. 

As long as $\lambda\neq 0$, we see that $C_\phi^{-2}$ is non-vanishing even in the limit $m_\phi \rightarrow 0$, which will impose an upper limit on the binding energy. Physically, we can interpret this as coming from the effective mass for the scalar mediator in the nugget, which caps the strength of the binding force. Conversely, given $(n_{\sat}/\mxbar^3, \mxbar/m_X)$, one can solve for $C_\phi^{-2}$ to derive a maximum quartic coupling $\lambda_{\max}$:\footnote{With only a quartic term in the scalar potential, there are two independent equations relating five dimensionless parameters: $C_V^2$, $g_\phi^2 m_X^2/m_\phi^2$, $\lambda$, $k_F/m_X$, and $m_*/m_X$. One therefore needs to specify three of the parameters to be able to fully determine the nugget properties. In particular, with only $n_\text{sat}/\mxbar^3$ and $\mxbar/m_X$ specified, only two of the remaining three degrees of freedom are fixed. 
The parameter $\lambda_\text{max}$ is the maximum $\lambda$ allowed for a given $n_\text{sat}/\mxbar^3$ and $\mxbar/m_X$ (corresponding to $m_\phi \rightarrow 0$ in the parameter space where $\mxbar/m_X \ll 1$).} 
\begin{align}
  \lambda_{\max} \simeq \frac{1}{C_\phi^2}\simeq\frac{ 3 \pi^2  n_{\sat}}{\mxbar^3} \left(\frac{\mxbar}{m_X}\right)^4 \left[1 - {1 \over 2}\left( \frac{3 \pi^2 n_\sat}{\mxbar^3} \right)^{1/3} \right] \, \qquad (\text{when}~~{\bar m_X \over m_X} \lesssim 1/2).
\label{eq:lambda_max}
\end{align}
We see that for small $n_\sat/ \mxbar^3$ and $\mxbar/m_X$, $\lambda_{\max}$ must be small as well. The requirement of a small quartic can also be understood intuitively: A small $n_\sat$ demands a large Fermi pressure, which forces us to consider relativistic constituents; this requires a large scalar mean-field to lower the effective mass. So the quartic coupling must remain small in order to keep $\mxbar$ small. 

\begin{figure}
{\centering \includegraphics[width=0.5 \textwidth]{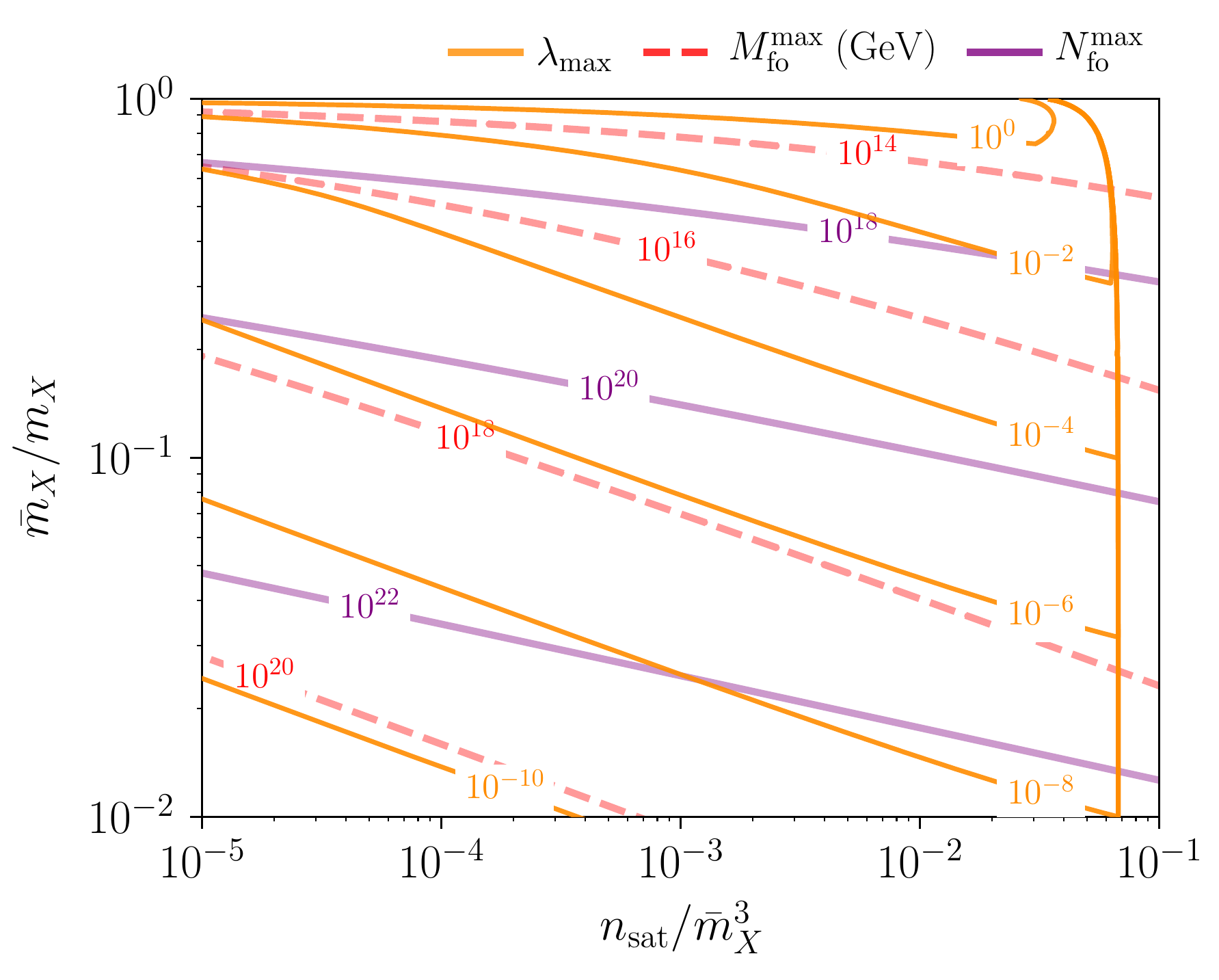}
}
\caption{Available physical parameter space $(n_{\rm sat}/\mxbar^3, \mxbar/m_X )$ for an extended nuclear model as described in \Sec{sec:extended_nuclear} with $V(\phi)={g_\phi^4 \over 3 \pi^2} \lambda \phi^4$. The orange boundaries indicate the maximal allowed value for $\lambda$, which indicates fine-tuning. [See 
Eqs.~\eqref{eq: CV Cphi definitions}, \eqref{eq:lambda_max} and surrounding discussion.] 
The dashed red (solid purple) shows the maximum achievable typical freeze-out nugget mass (number of constituents) $M_{\rm fo}$ ($\kfo$) given ${\SIDM \lesssim 1 \text{cm}^2/\text{g}}$ in the late Universe. [See Eqs.~\eqref{eq:Mfomax}-\eqref{eq:Nfomax}.] The synthesis temperature is (conservatively) taken to be $T_\syn = (m_X - \mxbar)/15$. A typical model is expected to have lower synthesis temperature and to require lower $\SIDM$ to accord with galactic structure observations and therefore smaller nugget freeze-out sizes and masses.
}
\label{fig:nuclear_param}
\end{figure}

\begin{figure}
{\centering
\includegraphics[width=0.8\textwidth]{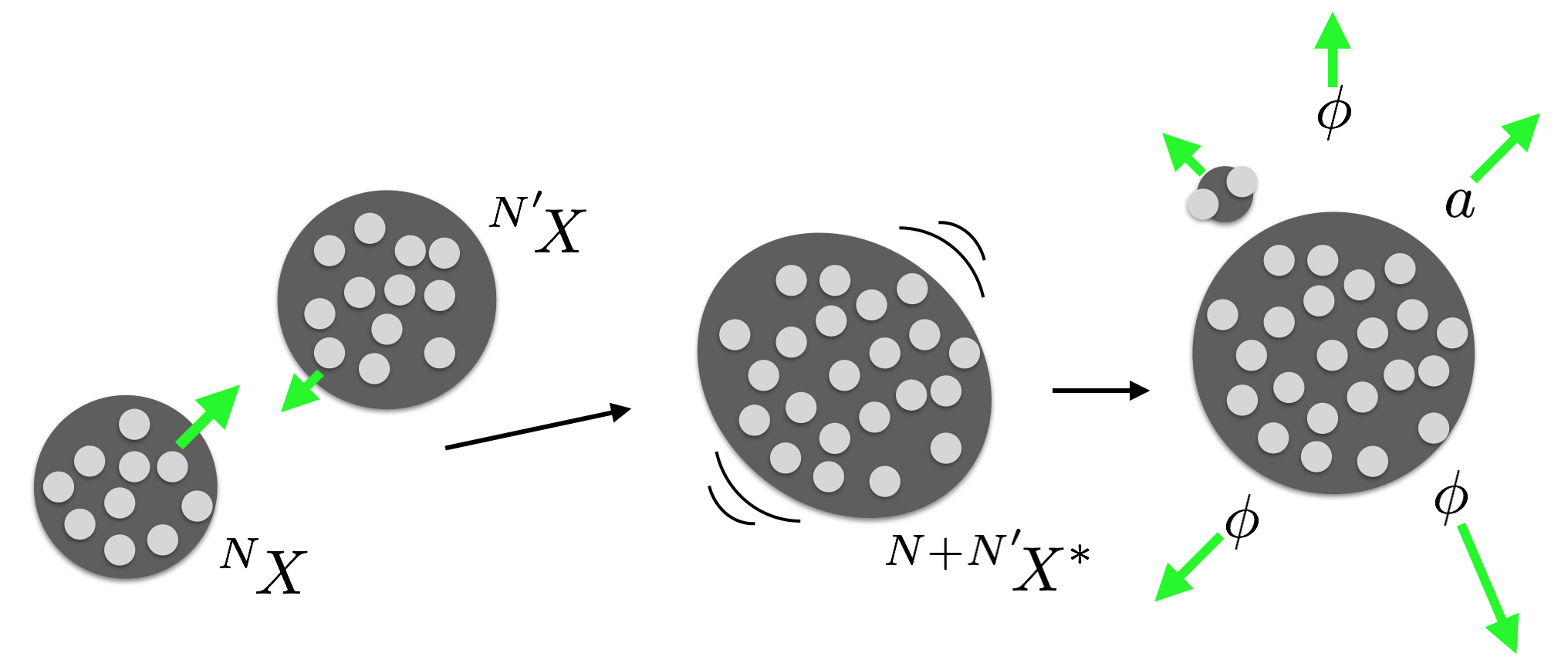}}\\

\caption{Schematic picture of large nugget fusion in the Compound Nucleus (CN) model. Cold fusion is possible due to the absence of a Coulomb barrier, and fusion remains exothermic up to arbitrarily large $N$.  {\bf 
In the early Universe}, synthesis begins at temperature $T_\text{syn} \lesssim {}^{2\!}X$ binding energy $
\times$ Boltzmann factor, and---absent a bottleneck at small $N$---proceeds to fusion processes as depicted above until reaching freeze-out due to number density depletion at typical size $N_\text{fo} \sim \left[{n_X \over H} \pi \left(4 \pi n_\text{sat} \over 3 \right)^{-2/3} \sqrt{T \over \bar m_X}\right]_{T = T_\text{syn}}^{6/5}$. {\bf In the late Universe}, ${\sigma_\text{DM} \over m_\text{DM}} \simeq \pi \left(N_\text{fo} \over {4 \pi \over 3} n_\text{sat} \right)^{2/3}{1 \over N_\text{fo} \bar m_X} $. SIDM bounds translate to upper bounds on $N_\text{fo}$, and there are stronger bounds from indirect detection if the fusion byproducts decay to SM particles. Furthermore, since the energy carried off by fusion byproducts is generally not redeposited, fusion is a cooling mechanism that can lead to accelerated core collapse at the centers of galaxies or collapse of (rare) early protohalos to form primordial black holes or exotic compact stars. Refer to \Fig{fig: saturated nugget pic} for definitions of parameters.}\label{fig: fusion picture}
\end{figure}

Connecting \Eq{eq:lambda_max} to early Universe synthesis, as we will see below in \Sec{subsec:synthesis}, fusion of large nuggets generally requires small $n_\sat$ (leading to larger cross sections) and/or $\mxbar$ (leading to larger number density). For a fixed $\lambda_{\max}$, 
the largest nugget consistent with SIDM constraints is synthesized in the scalar only limit. This is illustrated in \Fig{fig:nuclear_param}, which shows the available physical parameter space $(n_{\sat}/\mxbar^3, \mxbar/m_X )$ for an extended nuclear model 
with a potential term $V(\phi) = {g_\phi^4 \over 3 \pi^2} \frac{\lambda \phi^4}{2}$. 
The synthesis temperature is taken to be $T_\syn = (m_X -\mxbar)/15$, which, as we discuss in \Sec{subsec:synthesis}, is a conservative upper bound. 
Models with generic coupling and modest hierarchy $m_{\phi, V} \lesssim m_X$ largely populate the upper right corner of the physical space. The solid orange lines indicate contours of $\lambda_{\max}$ from numerical calculations. We see that in order to populate the lower left corner, $\lambda$ needs to be very small. The right most vertical curve shows the boundary of the densest nugget possible, obtained when the vector is decoupled, $C_V \rightarrow 0$. The right most orange line sits at ${n_\text{sat} \over \bar m_X^3} = {1 \over 3 \pi^2}$ at small $\mxbar/m_X$ as expected in the ultrarelativistic limit, up until ${\bar m_X \over m_X} \gtrsim 1/2$, where our analytic formulae become less accurate. 

Below we will discuss the detailed dependence of synthesized size $M_{\rm fo}$ and $\kfo$ on model parameters. Our conclusion is that large synthesized sizes require small $n_{\sat} / \mxbar^3$ and $\mxbar/m_X$, which can only be achieved for a very flat scalar potential---i.e.~when $\lambda$ is tuned to very small values.

\subsection{Scattering and Fusion Cross Sections}
\label{subsec:cxn}
For nuggets that are synthesized up to saturation sizes in the early Universe, both the size of the synthesized nugget and the interaction cross section in a halo today are controlled by the geometric cross section.  Here we briefly justify this claim and summarize standard results from nuclear physics that we will utilize in the rest of the analysis.

For large nuggets deep in the saturation limit, the range of the binding force is much smaller than the geometric size of the nugget. 
Since the nugget constituents must be relatively strongly interacting to bind in the first place, whenever two nuggets physically overlap, an interaction will very likely occur. Closely following the discussion in \cite{0486668274}, below we show that both the fusion and elastic scattering cross sections of saturated nuggets should be of order the geometric cross section under this assumption. 

Consider the interaction of two nuggets with radii $R_1$ and $R_2$. In all of our considerations, the interaction will occur in the nonrelativistic limit. 
Given the strong interaction and large spatial extent of the nuggets, the scattering problem can be solved via the Schr\"{o}dinger equation for a potential with depth of the order $N(m_X-\mxbar)$ and width of order the size of the nuggets $R_1+R_2$.  
We consider a general expansion of the cross section in terms of partial waves. For an incoming wave with wave number $k =  \mu v$, where $v$ is the relative speed and $\mu$ the reduced mass of the initial state nuggets, the geometric constraints of the nuggets translate to dominance of angular momentum modes $l\lesssim k (R_1 + R_2)$ in scattering processes. In terms of the partial wave amplitude $\eta_l$, the scattering cross sections can be parameterized
\begin{align}
  \sigma_{\rm sc}\sim \frac{\pi}{k^2} \sum_{l\lesssim k (R_1 + R_2)} (2l+1)|1-\eta_l|^2.
  \label{eq:sccxn}
\end{align}
In the strongly interacting limit, all the non-scattered waves are absorbed in fusion processes, and the fusion cross section is given by
\begin{align}
  \sigma_{\rm fus}\sim \frac{\pi}{k^2} \sum_{l\lesssim k (R_1 + R_2)} (2l+1)(1-|\eta_l|^2).
    \label{eq:fuscxn}
\end{align}
Geometric cross sections $\sigma \sim \pi(R_1+R_2)^2$ are recovered when $k (R_1 +R_2) \gg 1$,
 and for $\eta_l$ independent of $l$. Noting that $|\eta_l| \leq 1$ due to unitarity, it is also immediately apparent that $\sigma_\text{sc} \geq \sigma_\text{fus}$ and that the maximal fusion cross section corresponds to $\sigma_\text{fus} = \sigma_\text{sc} = \pi (R_1+R_2)^2$. In any case, as long as $|\eta_l| \not\approx 1$, we expect the fusion and scattering cross sections to be of the same order. The details of the cross sections will depend on the specifics of $\eta_l$. In the following, we will only be interested in order of magnitude estimates, and taking $\sigma_{\rm sc}\sim\sigma_{\rm fus}\sim \pi (R_1+R_2)^2$ will be sufficient.

For very low relative speeds, where $1/k \gtrsim (R_1+R_2)$,  $l=0$ scattering will dominate and the geometric cross section could be a significant underestimate. For fusion of two nuggets of similar size $\sim N$, we expect any such enhancement to be irrelevant as long as $N \bar m_X v \gg \left( n_\text{sat} \over N \right)^{1/3}$; since ${n_\text{sat} \over \bar m_X^3} < {1 \over 3 \pi^2}$, the enhancement is irrelevant in our galaxy as long as $N\gtrsim 10^4$. However, the enhancement could be relevant when a small nugget $\DN[N_2]$ interacts with a large one, as will be the case in the presence of a bottleneck discussed in \Sec{sec:bottleneck}.  In this case the cross section can be approximated as (see Ch.~VIII of \cite{0486668274})  
\beq
\sigma_{\rm fus} \sim \pi(R_1+ 1/p)^2 {\cal T} ~~~ \mbox{with}~~~{\cal T} = \frac{4 p p'}{(p + p')^2},
\label{eq:fus}
\eeq
where we have taken $R_2 \ll R_1$ and $k \approx p$ with $p = \sqrt{E_2^2 - m_{N_2}^2}$ the small nugget momentum, and $p' = \sqrt{E_2^2 - (\mxbar N_2)^2}$ is its effective momentum once inside the larger nugget.

\subsection{Size of Synthesized Bounds States}
\label{subsec:synthesis}

In the absence of strong bottlenecks at low $N$,\footnote{We reserve the details of the strong bottleneck scenario for \Sec{sec:bottleneck}.} 
early Universe synthesis proceeds until the typical size of nuggets reaches \cite{Gresham:2017cvl,Hardy:2014mqa}
\beq
N_\text{fo} = \gamma^{6/5} \qquad \text{with} \qquad \gamma \sim \left[ {n_X  \over H} \; \pi \left({4 \pi n_\text{sat} \over 3}\right)^{-2/3}  \sqrt{T_X \over \bar m_X}\right]_{t_\text{syn}}
\label{eq:synsize}
\eeq  
where $n_X$ is the conserved dark matter number density, $H$ is the hubble parameter, $T_X$ is the dark matter sector temperature, and $t_\text{syn}$ is the time when synthesis begins which is set by the two-body bound state binding energy ($T_{\text{syn}} \sim \BE_2/{\cal O}(10)$).  Here, $(\gamma H)^{-1} \ll H^{-1}$ is the interaction time scale. The estimate is insensitive to initial conditions \cite{2010kvsp.book.....K} and is therefore self consistent if $N_\text{fo} > N_\text{sat}$ so that geometric cross sections apply toward the end of synthesis.  Since $n_X(T_\gamma)$ and $H(T_\gamma)$ are known in the era of interest, the typical size of nuggets depends only on $n_\sat$ and $\mxbar$ once $T_\text{syn}$ is specified. Taking $T_X \sim T_\gamma$ we find,
\begin{align}
  \kfo \simeq 10^{12} \left({g_*(T_\text{syn}) \over 10}\right)^{3/5} \bigg( \frac{1 \; \GeV}{\mxbar}\bigg)^{\frac{12}{5}}\bigg( \frac{ \mxbar^3}{n_{\sat}}\bigg)^{\frac{4}{5}}
\bigg(\frac{T_{\rm syn}}{\mxbar} \bigg)^{\frac{9}{5}}\,.  \label{eq:Nfo}
\end{align}

The estimate in \Eq{eq:Nfo}, is strictly valid when fusion results in at most two nuggets and $\sigma v$ scales homogeneously as a function of $N$ \cite{2010kvsp.book.....K,Hardy:2014mqa,Gresham:2017cvl}. For the minimal model we considered in \cite{Gresham:2017cvl}, we argued that fusion will generally result in a single nugget in the final state (coagulation) along with many radiated mediators. With two nuggets in the final state, we showed that the final distribution becomes slightly broadened around $\kfo$ relative to the coagulation case. In models with multiple small nugget fragments in fusion final states, as long as the fragments are much smaller than the typical size we still expect the estimate to hold approximately, with perhaps some broadening of the final distribution about $\kfo$.

Synthesis begins when the rate for dissociation of small-$N$ states drops below the formation rate. Thus $T_\text{syn}$ depends on the cross sections and binding energies of low-$N$ bound states, which further depend on model details that are separate from the large-$N$, saturated nugget descriptions. Given that we wish to constrain the maximum sizes and masses of nuggets, we will take the conservative bound 
\begin{align} 
T_\syn \sim \frac{\BE_2}{30} \lesssim \frac{m_X - \mxbar}{15}\,.
 \label{eq: tsyn bound}
\end{align}
We have assumed $\BE_2 \lesssim 2(m_X -\mxbar)$ since otherwise large nuggets will dissociate into $\DN[2]$ in the large-$N$ limit.  In loose binding models, where the binding energy per constituent of large-$N$ states, ${m_X - \bar m_X}$, is a small fraction of $m_X$, we expect synthesis to begin well after ADM freeze-out, $T_\text{ADM} \sim m_X/30$ (when the constituents $X$ will have just become nonrelativistic and the symmetric DM component will have just annihilated away). However for strongly bound models in which $\bar m_X \ll m_X$ our conservative bound on $T_\text{syn}$ butts up against this ADM freeze-out time. In realistic models---especially for strongly bound models---we expect $T_\syn$ to be typically much smaller, leading to smaller final nugget sizes.    In any case our restriction on $T_\text{syn}$ leads to an upper bound on $\kfo$ and $M_\text{fo}$ which we discuss in the next section.

Imposing the conservative constraint, $T_\text{syn} \lesssim (m_X - \bar m_X)/15$, in the limit when $\lambda_\text{max}$ is small so \Eq{eq:lambda_max} holds and ${n_\text{sat} \over \bar m_X^3} \approx {\lambda_\text{max} \over 3 \pi^2}\left({\bar m_X \over m_X}\right)^{-4} $, we find
\begin{multline}
  \kfo \lesssim 10^{11} \left({g_*(T_\text{syn}) \over 10}\right)^{3/5} \bigg( \frac{1 \; \GeV}{\mxbar}\bigg)^{\frac{12}{5}} \lambda_\text{max}^{-\frac{4}{5}}
\left({\bar m_X \over m_X} \right)^{7 \over 5}\bigg(1 - {\bar m_X \over m_X} \bigg)^{\frac{9}{5}} \\
\lesssim 10^{10} \left({g_*(T_\text{syn}) \over 10}\right)^{3/5} \bigg( \frac{1 \; \GeV}{\mxbar}\bigg)^{\frac{12}{5}} \lambda_\text{max}^{-\frac{4}{5}} \qquad (\lambda_\text{max} \ll 1)
\end{multline}
where the second inequality follows from maximizing $x^{7/5}(1-x)^{9/5}$ in the interval $0 < x < 1$.
With a set mass scale, $\bar m_X$, we see that $\kfo$ is directly limited by naturalness alone.

\subsection{Products of Fusion} \label{sec:fusion products}

Here we address fusion byproducts, as predicted by the Compound Nucleus (CN) model. This is critical  for understanding both heat loss through fusion relevant for galactic halo evolution and indirect detection constraints. We will find that fusion generally produces an abundance of either force mediators (analogous to photons) or small nugget fragments (analogous to neutrons or alpha particles).  The CN model allows us to predict both the number of these fusion byproducts, as well as their energy spectra.

The essential feature of the CN model is that when two nuggets interact, they rapidly thermalize into an excited compound nucleus, which then decays through thermal emissions. The cross section for any given fusion process with initial state $i$ and final state $f$ factorizes as
\beq
\sigma(i,f) = \sigma(i\rightarrow C^*) {\Gamma(C^*\rightarrow f) \over \Gamma(C^*)}
\eeq where $C^*$ denotes the compound state whose characteristics depend only on the total energy, number of DM constituents, and angular momentum of the initial state.
Assuming the CN has a large density of states that is only slightly perturbed by particle emissions, the partial decay widths into various final states can then be assumed to take the form of a thermal spectrum characterized by a temperature, $T$ \cite{PhysRev.52.295}. More specifically, the partial decay width of a CN of size $N$ into a another CN state of size $N-k$ and a small nugget fragment $\DN[k]$ or light mediator $\DN[0]$ is given by
\begin{align}
  \Gamma_k(E^*) = 
g_k \int  {d^3 \vec p_k \over (2 \pi)^3}  e^{-Q(|\vec p|)/T(E^*)} \sigma_{N-k,k} v , \label{eq: decay spectrum 1}
\end{align}
where $g_k$ are the degrees of freedom of the state $\DN[k]$, $Q$ is the heat release, $E^*$ is the excitation energy of the $\DN[N]^{(*)}$,  $\vec p$ is the momentum of the fusion product $\DN[k]$, and $\sigma_{N-k,k}(\vec p)$ is the cross section for $\DN[k]$ absorption by the $\DN[N-k]^{(*)}$ nucleus. The heat release $Q = E^*_N - E^*_{N-k}$ is the difference in the excitation energies of the compound nucleus before and after emitting an $\DN[k]$ fragment. The excitation energy is given by $E^*_N = M_N - M_{N}^{(0)}$, where  $M_{N}^{(0)}$ is the nugget mass in its ground state.  Thus we have $Q  =  \sqrt{M_k^2 + \vec p^2} - M_{N}^{(0)} + M_{N-k}^{(0)} \approx   \sqrt{M_k^2 + \vec p^2} - k \bar m_X $, where $k \bar m_X$ approximates the difference in the ground state masses of the $\DN[N]$ and $\DN[N-k]$ nuggets. In terms of the fragment's kinetic energy, $\epsilon  = \sqrt{M_k^2+\vec p^2} - M_k$, the spectrum takes the form,
\begin{align}
 {d \Gamma_k \over d \Omega d \epsilon} = 
{g_k e^{-(M_k-k \mxbar)/T} \sigma_{N-k,k} \over (2 \pi)^3}    e^{-\epsilon/T} \epsilon(\epsilon + 2 M_k) . \label{eq: decay spectrum}
\end{align}
For mediator emissions, say $\phi$ for instance, $k=0$ and $M_k - k\mxbar = m_\phi$, and the exponential is simply a Boltzmann factor for $\phi$. In the limit where $\sigma_{N-k,k}$ is independent of $\vec p$, the peak of the distribution is at ${\epsilon \over M_k} = {T \over M_k} + \sqrt{\left({1 \over 2}\right)^2 + {T \over M_k}} - {1 \over 2}$, so the fragment's kinetic energy is generally of order the temperature. The fusion byproducts are emitted nonrelativistically when $T \ll M_k$ and relativistically when $T \gg M_k$.  When $\sigma_{N-k,k}$ is $\vec p$-independent, \Eq{eq: decay spectrum} can be integrated:
\begin{align}
  \Gamma_k(E^*) = 
e^{-(M_k-k \bar m_X)/T} \sigma_{N-k,k} {g_k \over \pi^2} T^2 (M_k + T). \label{eq: decay rate}
\end{align}
We see that the emission spectrum depends exponentially on $T$, with a weighting factor due to phase space and a coupling-dependent $\sigma_{N-k, k}$.
One can estimate $T$ by modeling an excited nugget as a low-temperature Fermi gas, such that $T\sim \sqrt{ E^* \bar m_X / N}$.\footnote{Recall that the heat capacity, ${d E^* \over d T}$, of a fermi gas is proportional to $N T / \epsilon_F$ at low temperature, $T \ll \epsilon_F$, and note that $\epsilon_F = \bar m_X$, here.} 
 For fusion of two ground-state nuggets of size $\sim N/2$, the excitation energy is order $E^* \sim \epsilon_\text{surf} N^{2/3} + \bar m_X N v_\text{rel}^2$ so that $T \sim \bar m_X \sqrt{{\epsilon_\text{surf} \over \bar m_X} N^{-1/3} + v_\text{rel}^2}$. The degenerate Fermi gas estimate of temperature is valid only when $T \ll \bar m_X$. Based on nuclear matter and our explicit calculations in Ref.~\cite{Gresham:2017zqi}, we expect $\epsilon_{\rm surf}$ to be of order the binding energy per particle, $m_X - \bar m_X$. Thus as long as $m_X \lesssim N^{1/3}\mxbar$, and the fusing mother particles are nonrelativistic, $T \ll \bar m_X$ and our approximation remains valid. Together with $M_k > k \bar m_X$, we see that non-mediator fragments ($\DN[k]$ with $k \neq 0$) are generally emitted nonrelativistically. 
For weakly bound models, the binding energy term $M_k - k\mxbar \sim k(m_X -\mxbar)$ can be comparable to $T$, and a large variety of small $\DN[k]$ fragments can be emitted. For deeply bound models, one expects $M_k \gg k \mxbar \gg T$, and the exponential factor dominates nugget emissions. 
Emissions will typically be dominated by one or two decay byproducts corresponding to the minimum of $M_k - k \bar m_X$. The minimum could occur for mediators (denoted by $\DN[0]$); requiring that synthesis can begin with ${X + X \rightarrow \DN[2] + \DN[0]}$ implies $m_{\DN[0]} <  2 m_X - m_2 < 2(m_X - \bar m_X)$. Thus, mediator emissions will likely dominate unless $m_{\DN[0]}$ is very near its maximum value, in which case emissions of $\DN[2]$ may be significant as well.

Our conclusions about the qualitative picture for fusion byproducs are summarized in Table~\ref{table:fusion byproducts}.   In \ref{sec:indirect detection} we will set limits on models in which fusion byproducts include mediators that can decay to SM states---potentially models in all four quadrants of the table. The limits have a mild dependence on whether the mediators are emitted relativistically or not.

\begin{table}
\noindent 
\makegapedcells
\setlength\tabcolsep{8pt}
\begin{tabular}{| p{1.3in} | p{2.3 in} | p{2.3 in} | }
\hline
\noindent\parbox[c]{\hsize}{} & $\textrm{\bf weak binding}$ (${{m_X - \mxbar} \ll \mxbar}$) & $\textrm{\bf strong binding}$ (${{m_X - \mxbar} \gtrsim \mxbar}$)\\
 \hline
 $\textrm{\bf heavy mediator}$ ($m_{\DN[0]} \sim 2 m_X - M_2$) & CN largely decays through emission of small $\DN[k]$, similar to neutron and $\alpha$ emissions for SM nuclei. Highly excited CN can decay into many low-$k$ states, which are emitted nonrelativistically. Mediator emissions can also be important if the coupling is large. &
CN decay is exponentially suppressed, as there is typically not enough energy locally to emit mediators or induce fragmentation. The detailed spectrum depends on mediator masses and binding energies; emission of the single particle species with minimum $M_k - k \bar m_X$ will strongly dominate.
\\
 \hline
 $\textrm{\bf light mediator}$ ($m_{\DN[0]} \ll 2 m_X - M_2$)  & 
Mediators will be readily emitted, although small $\DN[k]$ emissions may contribute significantly as well. Mediators may be emitted relativistically or nonrelativistically depending on the CN temperature. 
 & Emission of the lightest mediator strongly dominates; they can be relativistic  or nonrelativistic, depending on the CN temperature. \\
 \hline
\end{tabular}
\caption{Based on the compound nucleus (CN) model, a summary of expected fusion byproducts in formation of $\DN[N]$, according to the lightest mediator mass, $m_{\DN[0]}$, free-$X$ mass $m_X$, 2-body bound state mass, $M_2$, and average mass per constituent of saturated nuggets, $\bar m_X$. Emitted small nuggets (nugget fragments, $\DN[k]$ with $1 \leq k \ll N$) are generally nonrelativistic. This picture assumes a nonrelativistic initial state and that cross sections for small nugget fragment or mediator capture on a large nugget do not depend strongly on the identity or momentum of the fragment or mediator being captured.} \label{table:fusion byproducts}
\end{table}

\section{Nugget Interactions and the Structure of the Milky Way Galaxy}
\label{sec:interaction}

In the late Universe, the same interactions that lead to early nugget synthesis will also lead to dark matter self-interactions. These interactions can alter halo structures and possibly lead to indirect detection signals. Because interaction rates scale as ${\rho_\text{DM}  \sigma_\text{DM} v \over m_\text{DM}}$, and since $\rho_\text{DM}$ and $v$ are determined by observations, self-interactions are generally parameterized by $\SIDM$, which for a nugget with geometric cross sections and characteristic nugget number $\kfo$, is given by
\begin{align}
  \frac{\sigma_\DM}{m_\DM} \simeq
\pi \left(\frac{4\pi n_{\sat}}{3} \right)^{-\frac{2}{3}}
\frac{1}{\mxbar \kfo^{\frac{1}{3}}}
\,.
\label{eq:SIDM}
\end{align}
A sizable $\SIDM \sim 0.1 {\rm -} 1.0 \, \cm^2/g$ can soften dark matter cores and lead to better agreement with DM halo profiles \cite{0004-637X-547-2-574, 1538-4357-544-2-L87}. Additionally, nugget fusion is highly inelastic, such that one interaction per DM nugget in a halo lifetime can lead to contraction and potentially accelerated gravothermal collapse \cite{1968MNRAS.138..495L, RevModPhys.50.437}.  While understanding these effects in detail requires $N$-body simulations, we will address the effect of nugget fusions on central halo structure qualitatively in \Sec{sec:gravothermal}. Before that, in \Sec{sec:SIDM}, we examine the consequences of the conservative constraint, $\SIDM \lesssim 1 \, \cm^2/\text{g}$, coming from the bullet cluster \cite{Randall:2007ph} and galactic structure (see \cite{Tulin:2017ara} and references therein). 

\subsection{Self-Interaction Bounds for Nuggets}
\label{sec:SIDM}

At first glance \Eq{eq:SIDM} suggests that larger nuggets (with large $N_\text{fo}$) more easily evade a self-interaction constraint $\SIDM \lesssim 1 \, \cm^2/\text{g}$. However, as seen in \Eq{eq:Nfo}, $\kfo$ depends strongly on the nugget density and constituent mass. 
Taken together, we will see here that SIDM constraints actually put an \emph{upper} bound on $N_\text{fo}$ and $M_\text{fo}$.

More specifically, the SIDM bound $\SIDM \approx \pi \left( 4 \pi n_\text{sat} \over 3\right)^{-2/3}{\mxbar^{-1}N_\text{fo}^{-1/3}} \lesssim (\SIDM)_\text{max}$, 
effectively constrains the three-dimensional parameter space ($n_\text{sat},\bar m_X, T_\syn$) because $\kfo$ is itself a function of these three parameters. The constraint reads,
 \beq
\left(\frac{0.4 \text{MeV}}{ \bar m_X}\right)^{11/5} \left( { n_\text{sat} \over \bar m_X^3} \right)^{-2/5}  \left( {g_*(T_\syn) \over 10} \right)^{-1/5} \left( { T_\syn \over \bar m_X} \right)^{-3/5}  \lesssim  \left({ (\sigma_\text{DM}/m_\text{DM})_\text{max} \over \text{cm}^2/ \text{g} } \right).
\label{eq:general_constraint}
 \eeq
Holding the dimensionless parameters ${T_\syn \over \bar m_X}$ and ${n_\text{sat} \over \bar m_X^3} $ fixed, both $N_\text{fo}$ and $M_\text{fo}$ scale as negative powers of $\bar m_X$, which leads to upper bounds on both $N_\text{fo}$ and $M_\text{fo}$ as follows, 
\begin{align}
  \kfo &\lesssim
10^{20}\;
\bigg( {n_\sat \over \mxbar^3}\bigg)^{-\frac{4}{11}}
\bigg(\frac{g_*}{10} \bigg)^{\frac{9}{11}}
\bigg( \frac{T_\syn}{\mxbar}\bigg)^{\frac{27}{11}} \left({ (\sigma_\text{DM}/m_\text{DM})_\text{max} \over \text{cm}^2/ \text{g} } \right)^{12/11}
   \label{eq:Nfomax}\\
  M_{\rm fo} &\lesssim
10^{16} \; \GeV\;
\bigg( \frac{n_\sat}{\mxbar^3}\bigg)^{-\frac{6}{11}}
\bigg(\frac{g_*}{10} \bigg)^{\frac{8}{11}}
\bigg( \frac{T_\syn}{\mxbar}\bigg)^{\frac{24}{11}}  \left({ (\sigma_\text{DM}/m_\text{DM})_\text{max} \over \text{cm}^2/ \text{g} } \right)^{7/11}.
\label{eq:Mfomax}
\end{align}
The bounds on $M_{\rm fo}$ can be readily translated into an upper bound on the freeze-out nugget radius through $\pi R_\text{\rm fo}^2 \lesssim \left({\sigma_\text{DM} \over m_\text{DM}}\right)_\text{max} \; M_{\rm fo}^{\max}$. We have
\begin{align}
  R_{\rm fo} \lesssim 
1\,\mu {\rm m}
\bigg( \frac{n_\sat}{\mxbar^3}\bigg)^{-\frac{3}{11}}
\bigg(\frac{g_*}{10} \bigg)^{\frac{4}{11}}
\bigg( \frac{T_\syn}{\mxbar}\bigg)^{\frac{12}{11}}  
\left({ (\sigma_\text{DM}/m_\text{DM})_\text{max} \over \text{cm}^2/ \text{g} } \right)^{9/11}
\end{align}
These bounds are independent of the details of the nuclear model; they apply as long as large-large nugget fusions dominate near the end of synthesis and are described by geometric cross sections. 

The constraints can be relaxed if some nugget parameters exhibit large hierarchies: if $T_\syn \gg \mxbar$ or $n_\sat \ll \mxbar^3$. However, $T_\syn$ is bounded above as early Universe synthesis cannot occur when dissociation is efficient. We expect $T_\syn$ to be at least an order of magnitude smaller than the two-body bound state energy $\BE_2$, which must be smaller than $2(m_X -\mxbar)$ in order for large nuggets to be stable (c.f.~\Eq{eq: tsyn bound}). Substituting $T \lesssim (m_X -\mxbar)/15$ and $g_*\sim 10$ leads to the conservative bounds,
\begin{multline}
N_\text{fo} \lesssim 10^{17}  \left( n_\text{sat} \over \bar m_X^3 \right)^{-4/11}  \left(\left({\bar m_X \over m_X} \right)^{-1} -1 \right)^{27/11} \left({ (\sigma_\text{DM}/m_\text{DM})_\text{max} \over \text{cm}^2/ \text{g} } \right)^{12/11}
\\ \text{and} \qquad
M_\text{fo} \lesssim 10^{14} \;\GeV  \left( n_\text{sat} \over \bar m_X^3 \right)^{-6/11} \left(\left({\bar m_X \over m_X} \right)^{-1} -1 \right)^{24/11} \left({ (\sigma_\text{DM}/m_\text{DM})_\text{max} \over \text{cm}^2/ \text{g} } \right)^{7/11}.
\label{eq:Nfo_max_conserv}
\end{multline}
\Fig{fig:nuclear_param} shows the SIDM bounds on $\kfo$ and $M_\text{fo}$ in the ${n_\sat/\mxbar^3}\, , \, {\mxbar/ m_X}$ plane, taking $T_{\rm syn}\sim (m_X - \mxbar)/15$ but still accounting for the variation of $g_*$. 
 
\Eq{eq:Nfo_max_conserv} makes it clear that achieving sizes significantly larger than $\kfo \sim 10^{17}$ and $M_\text{fo} \sim 10^{14} \text{GeV}$ requires small $n_\sat/\mxbar^3$ and/or small $\mxbar/m_X$. However, our extended nuclear model reveals that achieving very small values for these dimensionless parameters is typically unnatural. This is shown by the orange contours in \Fig{fig:nuclear_param}, which indicate the maximum allowed value of quartic coupling, $\lambda$, required to achieve a given range of parameters $(n_\sat/\mxbar^3, \mxbar/m_X)$; alongside the corresponding maximum achievable $\kfo$ and $M_{\rm fo}$ contours (solid purple and dashed red contours, respectively), we see that achieving $\kfo \gg 10^{17}$ and/or $M_\text{fo} \gg 10^{14} \text{GeV}$ would require $\lambda \ll 1$. 
We expect a similar conclusion to hold for more general models with multiple flavors and additional terms in the scalar and vector interactions, as a small $n_\sat/\mxbar^3$ or $\mxbar/m_X$ is not protected by any specific symmetry. 

\begin{figure}
\includegraphics[width=0.45\textwidth]{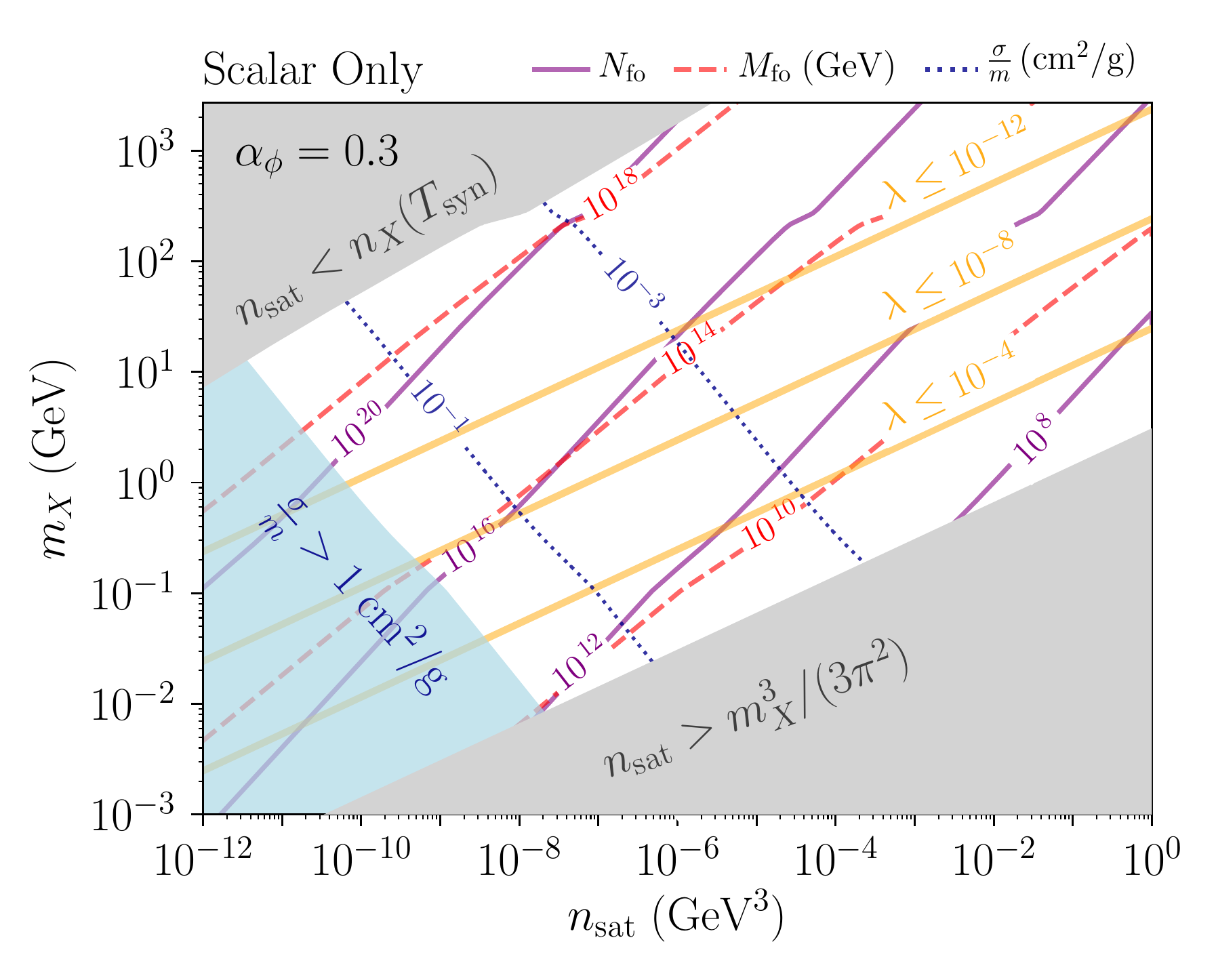} 
~~\includegraphics[width=0.45\textwidth]{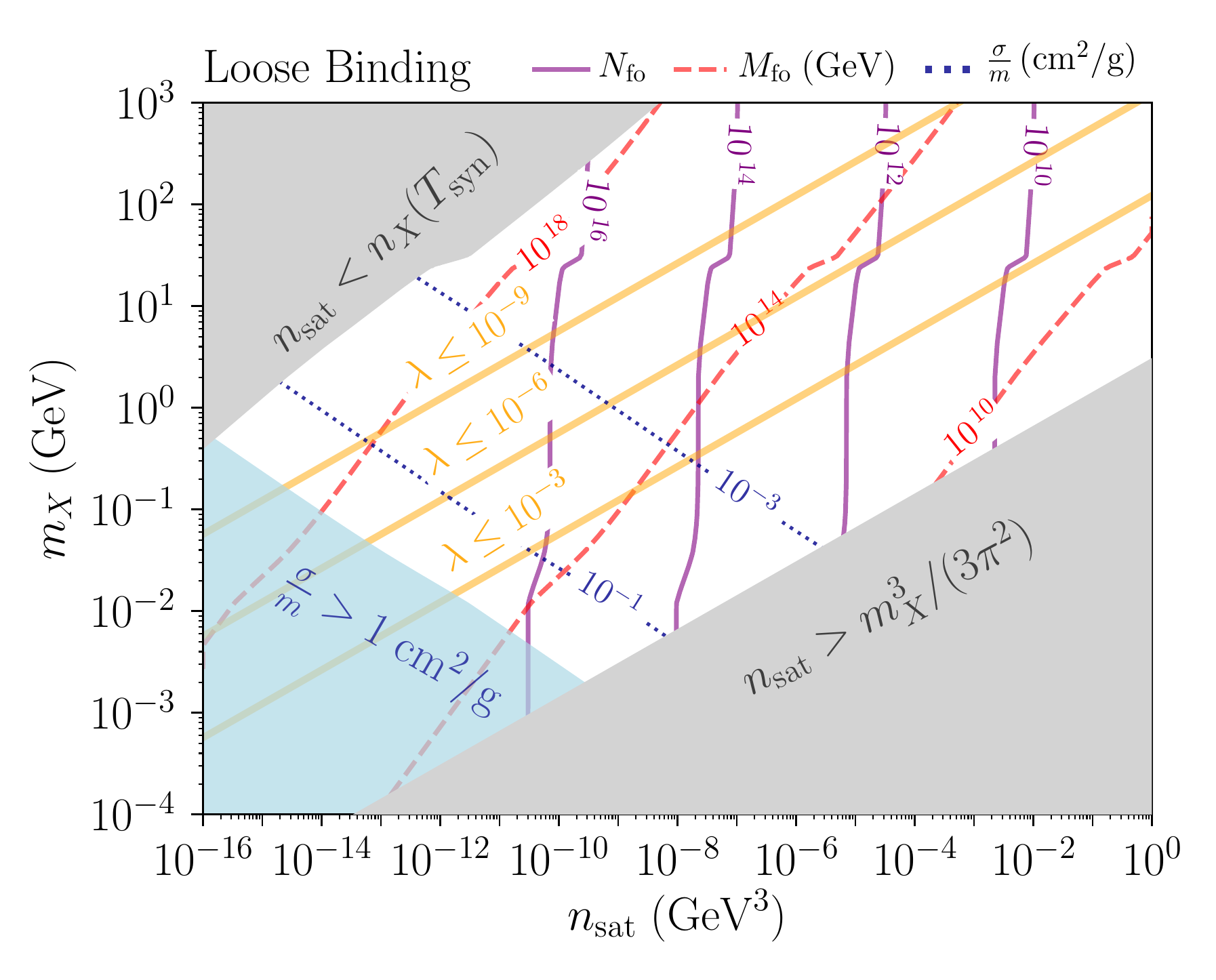}
\caption{Nugget size constraints for a scalar-binding model [left] and a benchmark loose-binding model ($\mxbar = 0.9~m_X$) [right], when no bottleneck is present. The dashed red (solid purple) contours indicate the characteristic mass (size) of the nuggets exiting synthesis, where $T_{\rm syn} = {\BE}_2/30 = \alpha_\phi^2 m_X/120$ in the left panel and $T_{\rm syn} = m_X/150$ in the right panel.  The light blue regions are excluded by the SIDM constraint $\SIDM < 1 \; \cm^2/g$. The dotted blue line shows contours of $\SIDM \sim 10^{-1}$ and $10^{-3}\;\cm^2/\gram$, indicating possible formation of a collapsed galaxy core. The solid orange curves indicate the maximum allowed scalar quartic coupling $\lambda$ for given $m_X$ and $n_\text{sat}$. In order to reach regions of larger $m_X$ at fixed $n_\text{sat}$, $\lambda$ must be progressively smaller.  Small $\lambda$ values may imply fine-tuning. [See Eqs.~\eqref{eq: CV Cphi definitions}, \eqref{eq:lambda_max} and surrounding discussion.] }
\label{fig:no_bottleneck}
\end{figure}

\Fig{fig:no_bottleneck} recasts these results in the $n_{\rm sat} - m_X$ plane for two different model extremes. The left plot corresponds to the scalar only limit, with $n_\sat /\mxbar^3 \simeq 1/(3\pi^2)$ and $T_\text{syn} \sim \BE_2/30 \approx \alpha_\phi^2 m_X /120$ with $\alpha_\phi = 0.3$.  The right figure corresponds to fixing $\mxbar = 0.9~m_X$, and choosing $\BE_2=2(m_X -\mxbar)$ so that $T_\syn = \BE_2/30 = m_X/150$.  This choice of the synthesis temperature is motivated as dissociation decouples typically at least a factor of $~30$ below the two-body binding energy. The blue regions are excluded by the SIDM constraint $\SIDM < 1 \;\cm^2/\gram$. The lower gray regions, where $m_X^3 \le 3\pi^2 n_\sat$, is a region of parameter space not realizable in an effective theory as defined in \Eq{eq:lagrangian}; c.f.~\Eq{eq: nugget scales}. The upper gray regions, where $n_\text{sat} \lesssim n_X(T_\text{syn})$, is a region where our model for synthesis would not apply; in particular, the model assumes that aggregation proceeds dominantly through 2-body interactions. The orange lines correspond to boundaries of the parameter space given a maximum $\lambda$. In both cases, a progressively smaller quartic is required to access regions with large nugget sizes and masses. For the scalar only case, $n_\text{sat}/\bar m_X^3 \approx (1/3 \pi^2)$ throughout most of the parameter space and the nugget size is largely controlled by $m_\phi$ (or the effective mediator mass inside the nugget), with efficient synthesis requiring strong binding with $\mxbar \ll m_X$. This leads to strong dependence on $m_X$ for both $\kfo$ and $M_{\rm fo}$. For the loose binding case, $\kfo$ depends on $m_X$ only through $g_*(T_\text{syn})$,\footnote{Note that \Eq{eq:Nfo} is equivalent to $ \kfo \simeq 10^{12} \left({g_*(T_\text{syn}) \over 10}\right)^{3/5} \left( \frac{ (1 \; \GeV)^3}{n_{\sat}}\right)^{\frac{4}{5}} \left(\frac{T_{\rm syn}}{\mxbar} \right)^{\frac{9}{5}}$.} leading to an almost $m_X$-independent contour for $N_\text{fo}$.

Additional constraints may be derived when additional model input is included. For instance, when the effective force range far exceeds the Bohr radius so that the 2-body interaction is effectively Coulombic, the binding energy is simply given by $\BE_2 \sim \alpha_2^2 m_X / 4$ when $\alpha_2 < 1$, and synthesis occurs roughly when $T_\syn \lesssim \BE_2 /30$ \cite{Gresham:2017cvl}. 
For synthesis to begin, the 2-body formation rate must exceed the Hubble rate at some point, which leads to $\alpha_2 \gtrsim 0.1 (m_X /100\, \GeV)^{1/3}$  \cite{Wise:2014jva}. Since the SIDM constraint \Eq{eq:general_constraint} puts a lower bound on the mass scale $m_X$ (or $\bar m_X$), this constraint along with $T_\syn \lesssim \alpha_2^2 m_X /120$ leads to the bound, 
\begin{align}
\alpha_2 \gtrsim 0.001 \left({g_* \over 10}\right)^{-{1 \over 17}} \bigg(\frac{\mxbar}{m_X} \bigg)^{-\frac{8}{17}} \bigg(\frac{n_\sat}{\mxbar^3} \bigg)^{-\frac{2}{17}}  \left({ (\sigma_\text{DM}/m_\text{DM})_\text{max} \over \text{cm}^2/ \text{g} } \right)^{-{5 \over 17}}\,.
\end{align}

We now turn to considering whether the highly inelastic and dissipative fusion interactions of nuggets require $\SIDM$ to be significantly less  than the naive SIDM limit $\SIDM \sim 1 \text{cm}^2 / \text{g}$ in order to remain consistent with observed galactic structure.

\subsection{Halo Core Gravothermal Collapse}
\label{sec:gravothermal}
A collisional gravitating system exhibits an instability as the core heats up and contracts, which leads to eventual gravothermal collapse \cite{1968MNRAS.138..495L, RevModPhys.50.437, Balberg:2002ue}. Most SIDM scenarios that have been studied include only simple elastic scattering processes. Nugget fusion processes, by contrast, are highly inelastic; both the binding energy and a significant fraction of the kinetic energy are lost to the fusion byroducts---usually light force mediators or small nugget fragments. As the core evolves toward gravothermal collapse, its density sharply increases, allowing all the hidden sector particles to thermalize. At the core boundary, the produced force mediators and nugget fragments are able to escape and dissipate heat. These byproducts may be reabsorbed by other nuggets. However, the absorption cross section is expected to be of the same order as the fusion reaction cross section, and the associated mean-free-path of these byproducts is typically very long compared to the core radius, 
\begin{align}
  \lambda_\text{mfp} \sim 0.4 \,{\rm Mpc}\, 
\bigg( \frac{0.4 \, \GeV/\cm^3}{\rho_\DM} \bigg)
\bigg( \frac{1\, \cm^2 /\gram}{\sigma_\DM/m_\DM}\bigg)
\end{align}
implying that nugget fusion is an extremely efficient energy loss mechanism. 

In the case of elastic SIDM, it was shown in \cite{Balberg:2002ue} that core collapse occurs soon after the core density is large enough such that the mean-free-path of the DM, $\lambda_\text{mfp} \sim m_\DM/(\rho_\DM \sigma_\DM)$, becomes shorter than the Jeans' length $\lambda_J \sim v/\sqrt{4\pi G \rho_\DM}$, where $v$ is the velocity dispersion. Numerical calculations showed that the time scale for this to occur is roughly given by 
\begin{align}
  t^{\rm elastic}_{\rm collapse} \sim \frac{240}{\rho^c_\DM v_c}\left(\frac{\sigma_\DM}{m_\DM} \right)^{-1}\,,
\label{eq:cooling}
\end{align}
where $\rho^c$ ($v_c$) is the central dark matter energy density (velocity dispersion) before the collapse. This time scale is an $\mathcal{O}(200)$ factor larger than the naive estimate $t_c \sim 1/[\rho_\DM^c v_c (\SIDM)]$.

By contrast with the elastic SIDM case, the nugget fusion case features an average loss of an ${\cal O}(1)$ fraction of the DM kinetic energy in each collision, which will result in infalling DM.  It has been shown in $N$-body simulations that inelastic processes can lead to enhancement of the central DM density \cite{Dubinski:1993df}. Since numerical analysis shows that increase in core density is a strong indicator of collapse, the collapse time with inelastic collisions is likely to be closer to, and perhaps even faster than,\footnote{We thank Haibo Yu for a discussion of this point in reference to their forthcoming work.} the naive estimate $t_c \sim 1/[\rho_\DM^c v_c (\SIDM)]$. For the Milky Way halo, and assuming an NFW profile up to the edge of the core, the cross section corresponding to a cooling rate of $10^{10}$ yrs roughly corresponds to $3 \times 10^{-2} \cm^2/\gram$ ($6\times 10^{-3} \cm^2/\gram$) at 1 kpc (0.1 kpc). 
  To illustrate when gravothermal collapse may be relevant, and for benchmark purposes, we show contours of $\SIDM \sim 0.1$ and $10^{-3}\, \cm^2/\gram$.  These smaller cross sections could possibly yield significantly different core structures than observed and may already be constrained.  A detailed analysis is reserved for future work.

\section{Other Nugget Constraints}
\label{sec:otherconstraints}

\subsection{Indirect Detection} \label{sec:indirect detection}
In fusion processes, as described by the Compound Nucleus (CN) model (see Sec~\ref{sec:fusion products}), many dark force mediators and/or nugget fragments (analogous to neutrons or alpha particles) may be emitted. The nugget fragments are stable due to conserved DM number, while the mediators may decay back to the SM. The decays can be mediated by couplings between the dark sector and the SM. For instance, the scalar may mix with the Higgs and the vector may kinetically mix with hypercharge. Decay of the DM mediator fusion byproducts into SM particles can lead to injection of energy into the cosmic microwave background (CMB) or excess photon flux from galaxies, mimicking the case of DM decay or annihilation. Here we discuss indirect detection constraints from the CMB and photon flux, in turn.

If charged particles or photons are produced in the decay, they can disrupt the CMB spectra after recombination. This places a constraint on the energy deposited into the hydrogen gas, which can be written as \cite{Giesen:2012rp, Slatyer:2015jla}
\begin{align}
\left[ f_{\rm eff} \left(\frac{\sigma_{\DM}v}{m_{\DM}}\right) \right]_{z\sim 600}
 \lesssim 10^{-14} \; {\cm}^2/{\gram}\,,
\end{align}
where $f_{\rm eff}$ is an efficiency factor that depends on the annihilation processes. For WIMPs that annihilate into gauge bosons or fermions, $f_{\rm eff}$ ranges from $\sim 0.1 - 0.5$. 
For nuggets, one expects $f_{\rm eff}$ to be significantly suppressed as fusion reactions only release a small fraction of the rest energy of the DM; namely, we expect
\begin{align}
 f_{\rm eff} \sim f_\gamma \frac{E^*}{2 \kfo \mxbar} \sim 
f_\gamma\frac{\epsilon_{\rm surf}}{2 \mxbar \kfo^{\frac{1}{3}}} \,,
\end{align}
where $f_\gamma$ is an $\mathcal{O}(1)$ efficiency factor proportional to the fraction of the released energy ejected into the CMB. Here $E^* \sim \epsilon_{\rm surf}\kfo^{2/3}$ is the excitation energy in the small velocity limit. The velocity here is small $v \sim \sqrt{T/(\kfo \mxbar)}$, \footnote{If the nuggets have fallen out of kinetic equilibrium, the velocity will be even smaller as it scales like $T/\mxbar$.} which leads to a further suppression of the constraint, and we have
\begin{align}
  \frac{\sigma_\DM}{m_\DM}
\lesssim 1 \; {\cm}^2/{\gram}\,
\sqrt{\frac{\mxbar^3}{\epsilon^2_{\rm surf} T}}
\bigg(\frac{0.1}{f_\gamma}\bigg)
\bigg(\frac{\kfo}{10^{16}} \bigg)^{\frac{5}{6}}\,.
\label{eq:CMB_constraint}
\end{align}
This bound can compete with the naive SIDM bound $\SIDM \lesssim 1 \,\cm^2/\gram$, depending on the ratio $\mxbar/\epsilon_{\rm surf}$. 

Much stronger constraints can be derived from the galactic photon flux.  Depending on the mass of the mediator, and whether it is emitted relativistically or nonrelativistically, X-ray or gamma ray constraints may dominate.  
To derive bounds on nugget fusion cross sections from decay of fusion byproducts within galaxies, here we will follow \cite{Essig:2013goa} which derives bounds on DM annihilation and decay rates. To be concrete, we consider only scalar mediators that primarily decay into $\mu^+\mu^-$ or $e^+ e^-$; constraints will only become stronger if decay to hadrons (including pions, which go directly to $\gamma \gamma$) is permitted. 
Other models involving DM vector mediators or other alternative decay channels can also be constrained, but we do not expect the constraints to be substantially different as compared to the scalar decay case. The incoming photon flux can be computed as
\begin{align}
  \frac{d\Phi_\gamma}{dE} = \frac{r_\odot}{8\pi}
\frac{\rho_\DM^2\langle \sigma v\rangle_{\rm DM}}{m^2_\DM}
N_\phi\frac{dN_\gamma}{dE}J\,,
\end{align}
where $r_\odot \sim 8.5$ kpc, $\rho_\DM \sim 0.3 \,\GeV/\cm^3$ is the local DM density, $N_\phi$ is the average number of scalar mediators emitted per fusion event, and $dN_\gamma/dE$ is the average differential energy spectrum for each emitted mediator.  The factor $J$ is an $\mathcal{O}(1-10)$ dimensionless number characterizing the squared density of the DM along the line-of-sight and solid angle for a given observation. We will ignore extragalactic contributions, which could, depending on the amount of substructure \cite{Liu:2016cnk}, significantly enhance the signal.  

The number of mediators emitted per fusion event, $N_\phi$, can be estimated using the CN model (see Sec~\ref{sec:fusion products}). The average energy of each emitted mediator is approximately $m_\phi + T^*$, where $T^*$ is the temperature of the excited CN. The CN's excitation energy and thus temperature decreases with each emission since $T^* \approx \sqrt{E^* \bar m_X / N_\text{fo}}$. If only $\phi$'s are emitted, we have ${d E^* \over d N_\phi} \approx - (m_\phi + T^*(E^*))$ so that 
\begin{align}
  N_\phi \sim \frac{E^*_0}{m_\phi} \left( {2 m_\phi \over T^*_0} \left(1- { \ln(1+ T^*_0 / m_\phi) \over {T^*_0/m_\phi}} \right)\right) \sim 
\frac{1}{\max \{ m_\phi, T^*_0 \}}
\left(\frac{\kfo \mxbar v^2_{\rm rel}}{4} + {\epsilon_{\rm surf}} \kfo^{\frac{2}{3}}.
\right)
\end{align} 
It is convenient to absorb the dependence on $N_\phi$ into a dimensionless factor $f_{\rm ind}$, defined as
\begin{align}
  f_{\rm ind} \equiv 
\frac{N_\phi \max \{ m_\phi, T^* \}}{m_\DM}
\sim 
\frac{v^2_{\rm rel}}{4} + \frac{\epsilon_{\rm surf}}{\mxbar} \kfo^{-\frac{1}{3}}\,.
\end{align}
The indirect constraint can then be rewritten as a bound on $f_{\rm ind}$ times $\SIDM$. Using $f_{\rm ind} \gtrsim v^2 \sim 10^{-6}$ then leads to a conservative upper bound on $\SIDM$.
\begin{figure}[ht]
\includegraphics[width=0.45\textwidth]{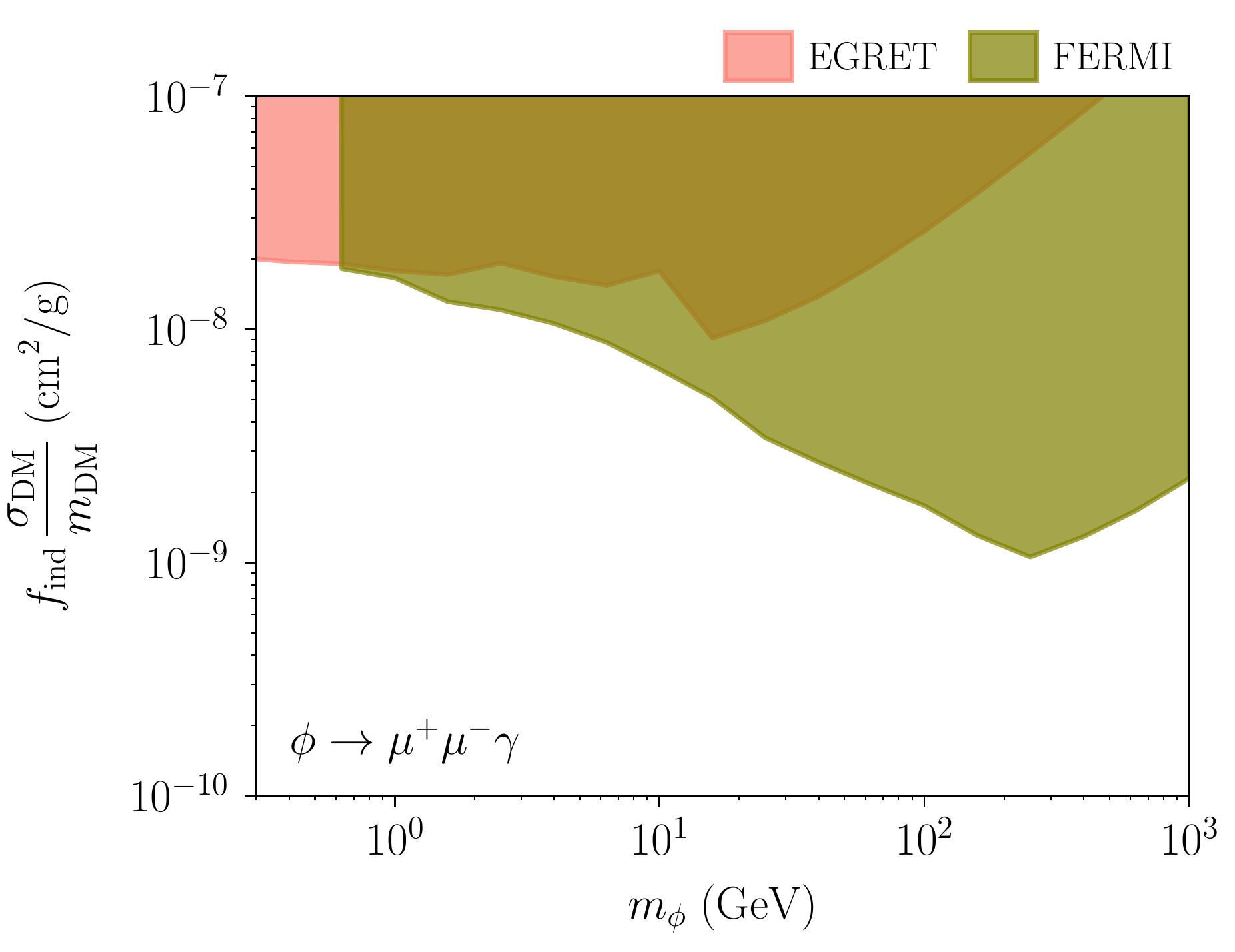} ~~\includegraphics[width=0.45\textwidth]{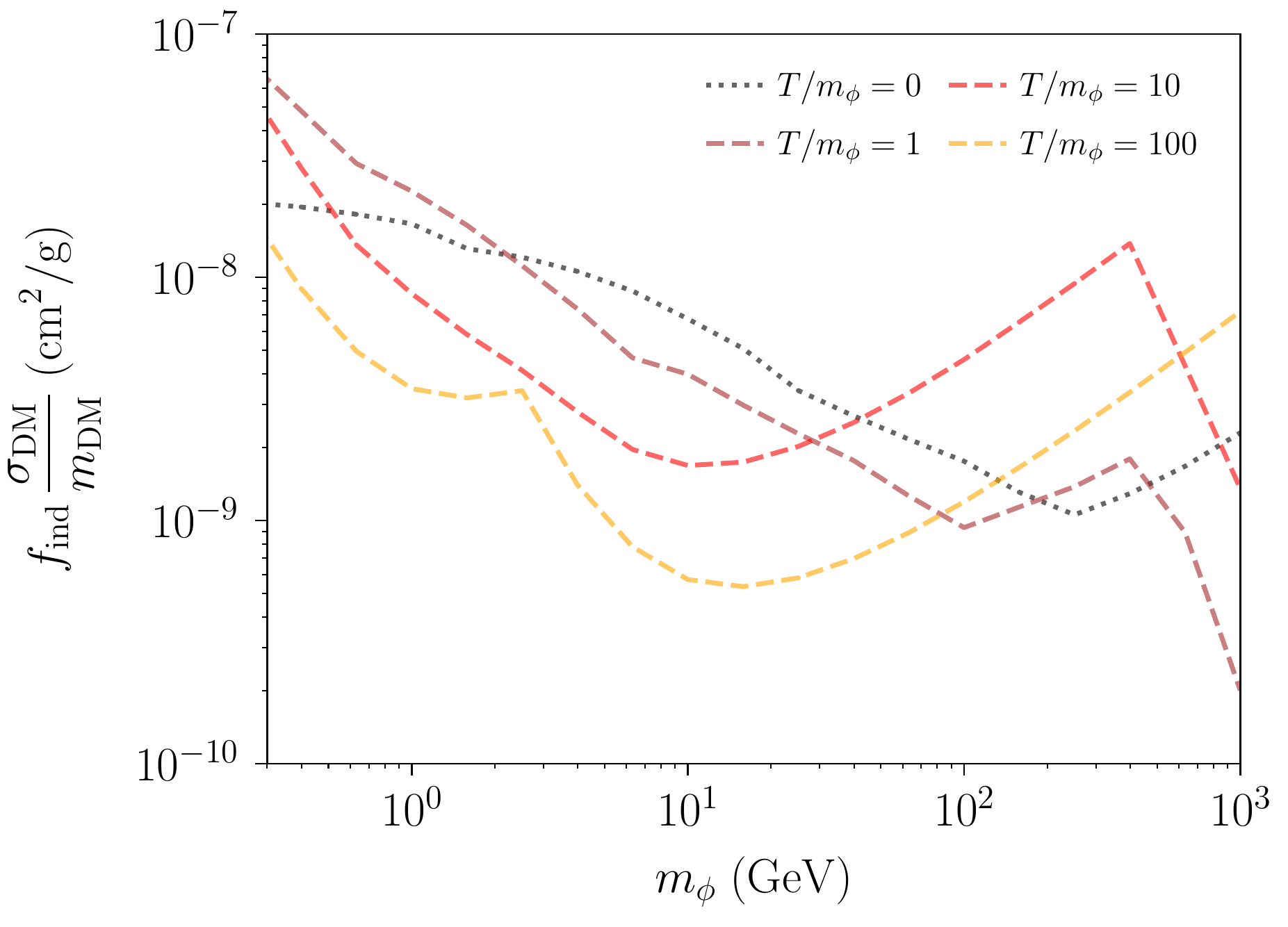}
\includegraphics[width=0.45\textwidth]{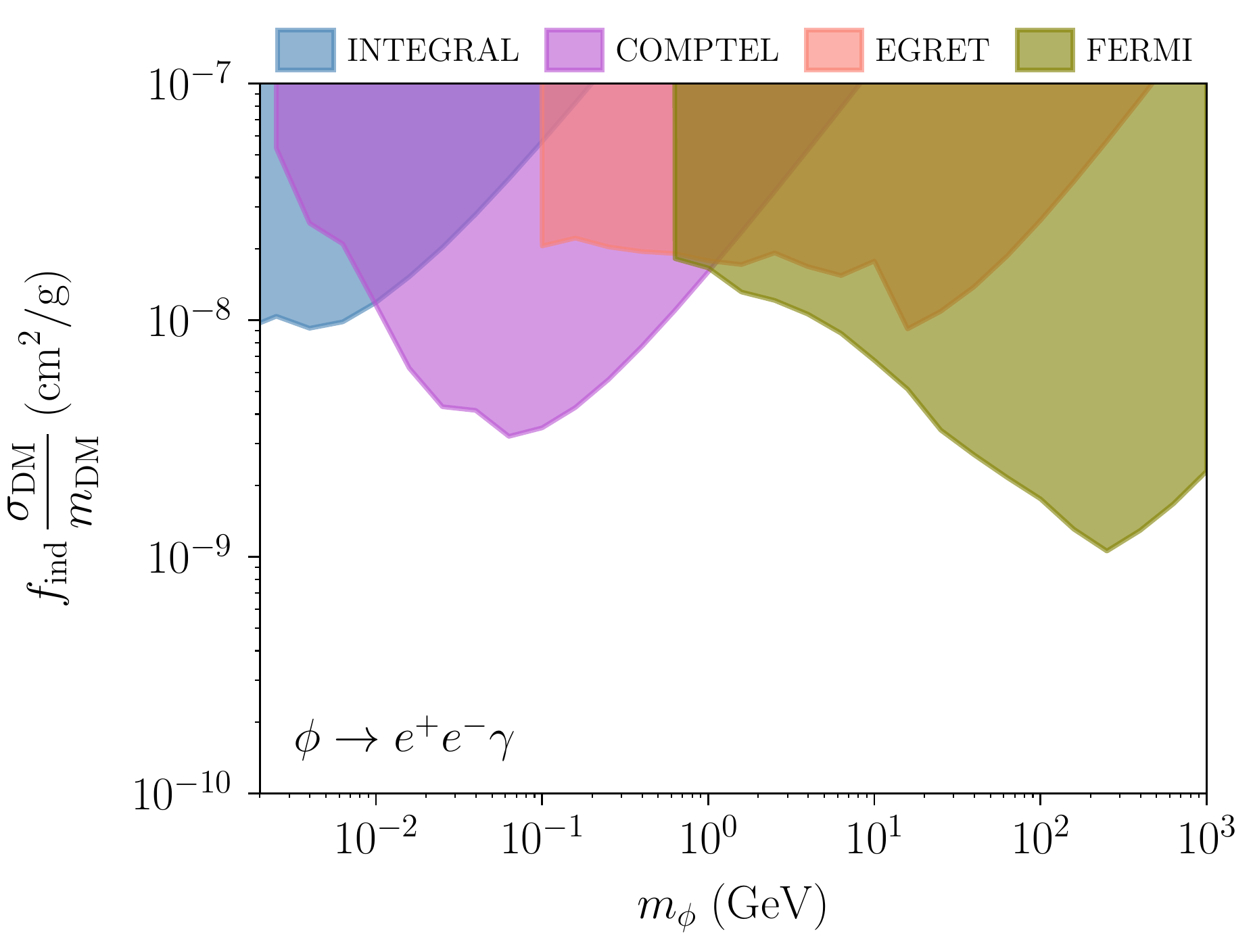} ~~\includegraphics[width=0.45\textwidth]{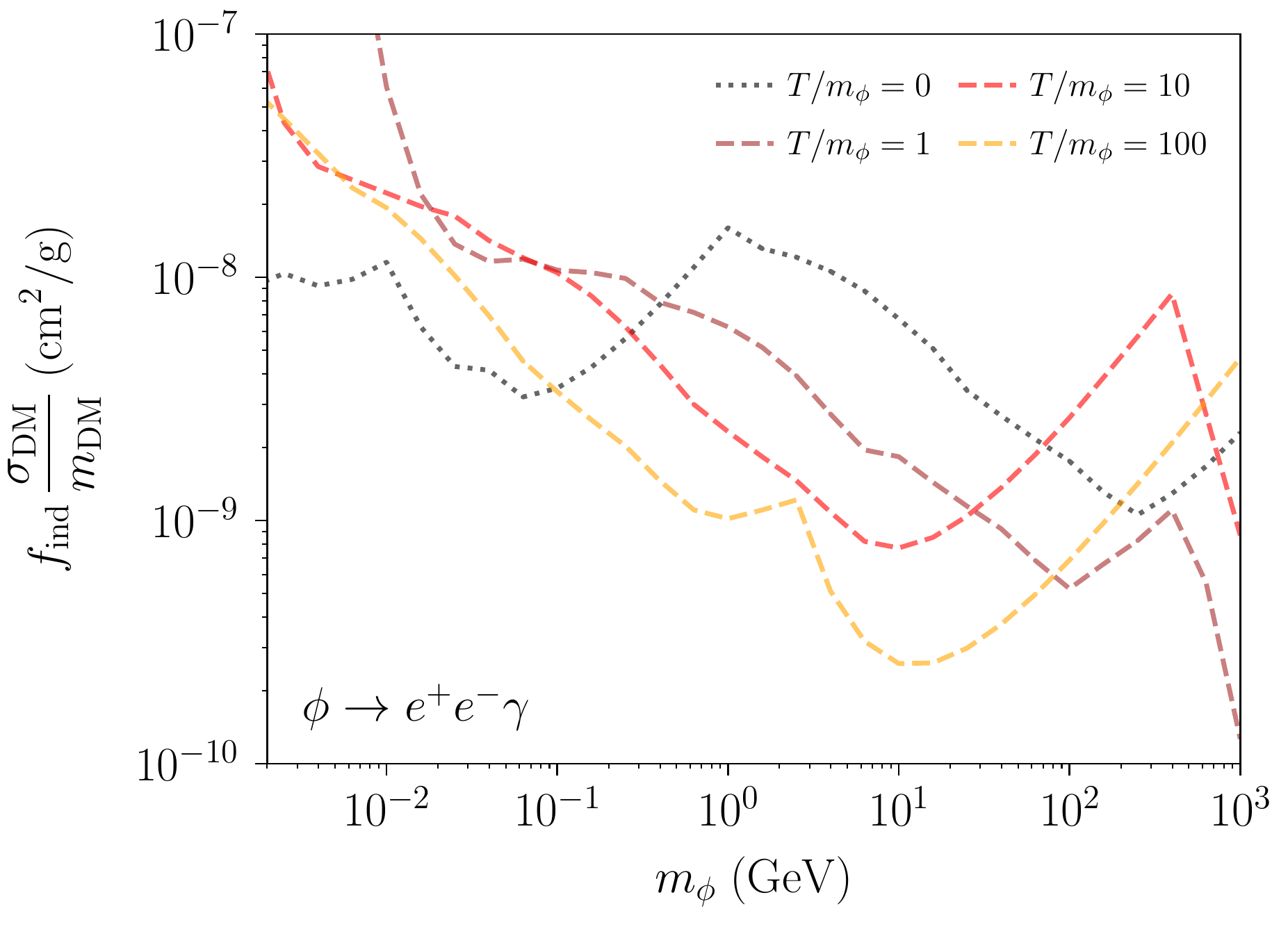}
\caption{Indirect detection constraints for a scalar mediator decaying into $\phi \rightarrow \mu^+ \mu^- \gamma$ (top) and $\phi \rightarrow e^+ e^- \gamma$ (bottom). \emph{Left}: Constraints for nonrelativistic emissions, where the emitted mediators are assumed to be at rest. \emph{Right}: Constraints for relativistic emissions, where the momentum of the mediator is assumed to follow a Boltzmann distribution.}\label{fig:indirect_limit}
\end{figure}

Following the calculations in \cite{Essig:2013goa}, we consider the $\gamma$ spectrum from soft collinear splitting of photons from the charged decay products of a scalar mediator. Conservatively, we have ignored additional flux that can come from hard bremsstrahlung and muon decay, which could only tighten the bounds. Here we focus on two topologies: $\phi \rightarrow \mu^+\mu^- \gamma$ and  $\phi \rightarrow e^+e^- \gamma$. For a mediator decaying from rest, 
\begin{align}
  \frac{dN_\gamma}{dE} \simeq \frac{2\alpha_{\rm EM}}{\pi E_\gamma}
\left\{1-\frac{2E_\gamma}{m_\phi}+
\left(1-\frac{2E_\gamma}{m_\phi}+ \frac{2E^2_\gamma}{m^2_\phi}\right)
\log\left[\frac{m_\phi^2}{m_l^2}\left(1- \frac{2E_\gamma}{m_\phi}\right) \right]
 \right\}\,,
\label{eq:gamma_spectrum}
\end{align}
where $m_l = m_{e,\mu}$ and the spectrum is assumed to be zero when the logarithm goes to zero at large enough $E_\gamma$. For nonrelativistic emissions ($T^* \ll m_\phi$), \Eq{eq:gamma_spectrum} gives the differential photon spectrum. For relativistic emissions ($T^* \gg m_\phi$), the photon spectrum is approximated by
\begin{align}
  \frac{dN_\gamma}{dE} \sim 
\frac{\int d\Omega\, d^3 \vec v\, e^{-\frac{m_\phi}{\sqrt{1-v^2}T^*}}\left(\frac{dN_\gamma}{dE d\Omega}\right)_{\textrm{$\vec v$-boosted}}}{\int d^3 \vec v\, e^{-\frac{m_\phi}{\sqrt{1-v^2}T^*}}}\,,
\end{align}
where we have boosted \Eq{eq:gamma_spectrum} according to the $\phi$ emission spectrum as estimated by the CN model. 
\Fig{fig:indirect_limit} shows the constraints on $f_{\rm ind}(\SIDM)$ for the decay channel $\phi \rightarrow \mu^+\mu^- \gamma$ (top) and $\phi \rightarrow e^+e^- \gamma$ (bottom). The left panels show the constraints for nonrelativistic emissions, where the spectrum is given by \Eq{eq:gamma_spectrum}. The right panels show the constraints for relativistic emissions with different $T^*$, combining all experiments. These constraints are generally stronger than SIDM bounds, and perhaps comparable to a gravothermal collapse bound discussed in \Sec{sec:gravothermal}, which will translate to similar limits on $\mxbar$ and $\kfo$ through Eqs.~(\ref{eq:general_constraint})-(\ref{eq:Nfomax}).  

Now we apply indirect detection constraints to the scalar-only model studied in \cite{Wise:2014jva}. Here, all the nugget properties can be explicitly computed from Lagrangian parameters \cite{Gresham:2017zqi}, and we take $T_{\syn}\sim \alpha_\phi^2 m_X/120$. Since the nuggets are deeply bound in this regime, the binding energy dominates over the kinetic energy in fusion reactions, and $f_{\rm ind} \sim \epsilon_{\rm surf} \kfo^{-1/3}/\mxbar$. The excited nugget is also expected to have very low excitation temperature where nonrelativistic emissions dominate. Then using Eqs.~\eqref{eq:Nfo} and \eqref{eq:SIDM} with ${n_\text{sat} \over \bar m_X^3} \approx {1 \over 3 \pi^2}$ (see \Eq{eq: nugget params} and \Fig{fig:nuclear_param}), the indirect detection constraint can be rewritten as
\beq
\left({\epsilon_\text{surf}/m_X \over 10} \right)  \left( 10 \over g_*(T_\text{syn}) \right)^{2/5} \left( m_X \over \mxbar \right)^{6/5} \left( 100\, \text{GeV} \over m_X  \right)^{7/5}  \left(0.1 \over  \alpha_\phi \right)^{12/5}  \lesssim { \left( f_{\rm ind} {\sigma_\text{DM} / m_\text{DM}}\right)_\text{max} \over 10^{-9}\, \cm^2/{\gram} } .
\eeq

Keep in mind that binding requires ${\mxbar \over m_X} \approx \left({3 \pi \over 2 \alpha_\phi }{m_\phi^2 \over m_X^2}\right)^{1/4} < 1$.  And for synthesis to begin and proceed efficiently in the early Universe, one requires ${\alpha_\phi \over 0.1} \gtrsim \left({m_X \over 100 \text{GeV}}\right)^{1/3}$ and $m_\phi < \BE_2 = {\alpha_\phi^2 m_X \over 4}$, respectively \cite{Wise:2014jva}. This last condition implies ${\mxbar \over m_X} \lesssim \alpha_\phi^{3/4}$. In a model where $\phi$ decays primarily to muons, for example, at the most generous the constraint is $\left( f_{\rm ind} {\sigma_\text{DM} / m_\text{DM}}\right)_\text{max} \sim 10^{-7}$. Satisfying the constraint along with the conditions just mentioned requires $\alpha_\phi \gtrsim 0.1$ and $m_X \gtrsim 20\,\GeV$; though note that as $\alpha_\phi$ becomes nonperturbative, the estimate for $\BE_2$ (and thus $T_\text{syn}$) and the $\DN[2]$ formation rates that fed into the ${\alpha_\phi \over 0.1} \gtrsim \left({m_X \over 100 \,\text{GeV}}\right)^{1/3}$ condition break down. Overall, our constraints are competitive with those studied in \cite{Wise:2014jva}, and a scalar-only model with moderate $m_X$ and $\alpha_\phi$ can still be viable.

\subsection{Cooling in Early Protohalos} \label{sec: proto halo cooling}

We now consider dark star formation in the early Universe, through cooling of smaller protohalos that virialize and break away from the Hubble flow at high redshifts. In contrast to the SM where the Coulomb force provides a means to dissipate energy to form a disk, which then fragments to form stars, one expects dark star formation to proceed directly through the highly efficient and exothermic fusion processes in these protohalos.  Note that this is also in contrast to models where dark star formation has been considered in the presence of a dark Coulomb force ({\em e.g.} \cite{Fan:2013tia,Agrawal:2017pnb,Foot:2014uba, Buckley:2017ttd}).  We show that at the beginning of structure formation, if nuggets are the primary DM component, an SIDM bound not too much stricter than $\SIDM \sim 1\, \cm^2 / \gram$  allows for only very rare protohalos to have completely collapsed to stars due to cooling through fusion. 

Within a model of bottom-up hierarchical structure formation (see {\em e.g.}~\cite{Barkana:2000fd,Ripamonti:2005ri} for a review), up to corrections of order $\Omega_m(z)-1$, the density of overdense regions relaxes to $\rho_{\rm coll}(z)  \sim 18\pi^2 \rho_{\rm crit}(z)$ after breaking away from the Hubble flow and virializing, where $\rho_{\rm crit}(z) \sim \rho_{\rm crit}(0) \Omega_m (1+z)^3$  is the critical density at the redshift $z$ of the collapse. The velocity dispersion at the virial radius is given by 
\begin{align}
  v_{\rm dis} \sim \sqrt{{3 \over 5}{ G M \over R}} \sim 3 \times 10^{-3} \;
\sqrt{1+z}
\bigg( \frac{M}{10^{15}\; M_{\odot}}\bigg)^{\frac{1}{3}}
\end{align}
where $M$ is the halo mass. The cooling timescale can be estimated as $(\rho v {\sigma_\text{DM} \over m_\text{DM}})^{-1}$, which needs to be {\em at least} less than $H_0^{-1}$ for gravothermal collapse to be relevant. The proper timescale should in fact be somewhat lower as we have not included here the effects of tidal stripping on star formation.  Even with this generous formation time allowance, we will find that only very rare protohalos can form dark stars.   With this requirement, the cross section for a protohalo to form interesting structure is thus
\begin{align}
  \frac{\sigma_\text{DM}}{m_\DM}
\gtrsim
50 \; \cm^2 /\gram \;
(1+z)^{-\frac{7}{2}}
\bigg( \frac{10^{15}\; M_{\odot}}{M}\bigg)^{\frac{1}{3}}\,.
\label{eq:SIDM_protohalo}
\end{align}

\begin{figure}[ht!]
\includegraphics[width=0.6 \textwidth]{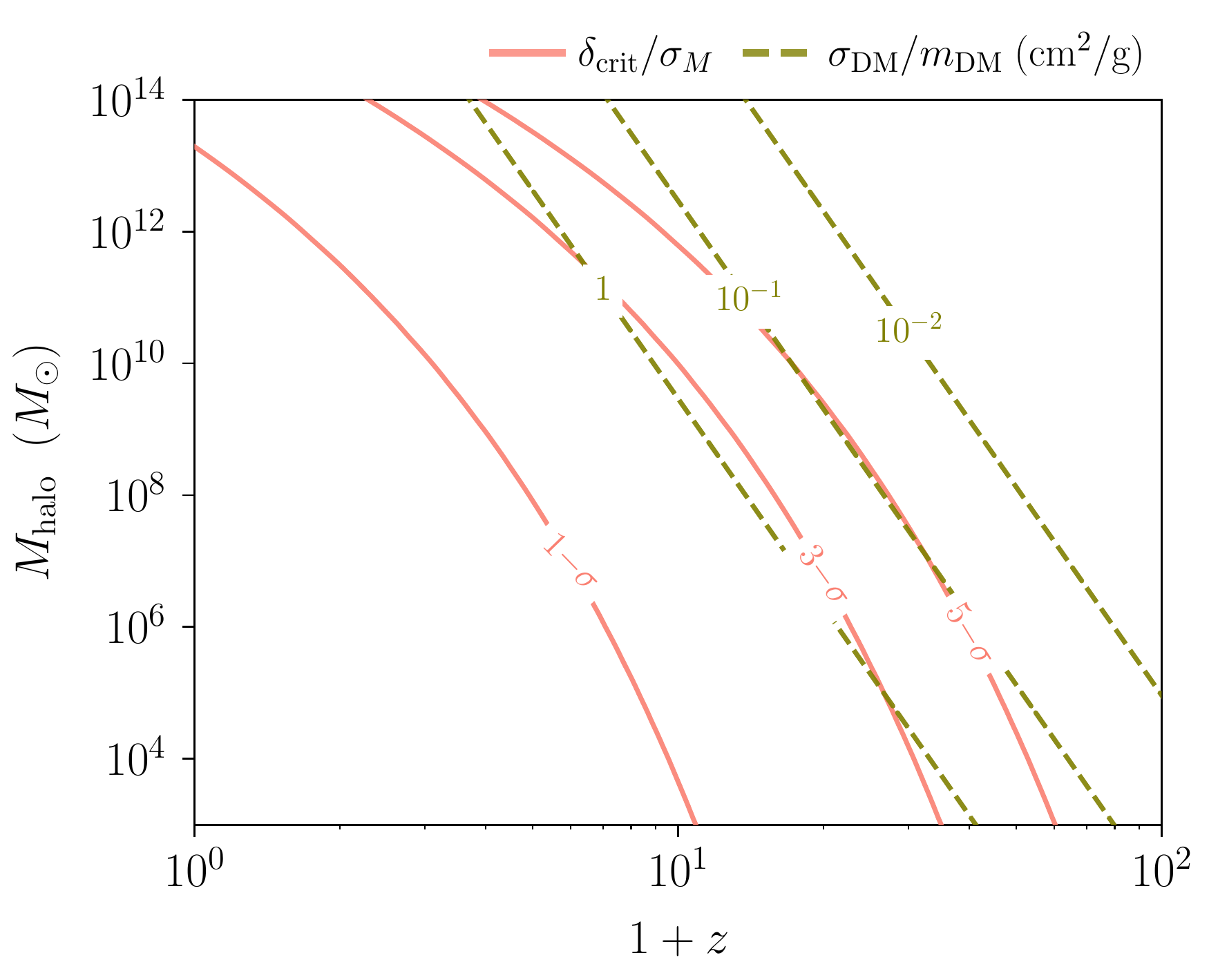}
\caption{Solid Red: contours of the halo mass ($M_{\rm halo}$) vs red-shift ($1+z$) that correspond to $n$-$\sigma$ fluctuations in the Press-Schechter model. Dashed Green: regions with cooling timescale of order $H_0^{-1}$ assuming $\SIDM \sim (1,10^{-1},10^{-2}) \; \cm^2/\gram$}
\label{fig:protohalo}
\end{figure}

In the Press-Schechter model, 
regions collapse and virialize roughly when the linear density perturbation smoothed over spherical regions with mass scale, $M$, modeled as a Gaussian random field with $M$-dependent variance $\sigma^2$, fluctuates above a certain $z$-dependent critical value. So a halo of given mass collapsing at redshift $z$ corresponds to a certain number of standard deviations, $\sigma$, fluctuation. 
\Fig{fig:protohalo} shows the $\sigma$ contours (solid red) in the halo mass versus redshift plane.\footnote{The contours were digitized from Fig.~6 of \cite{Barkana:2000fd}. The assumed $\sigma^2(M)$ spectrum is from \cite{Hu:1998tj}.} The dashed green line shows the required $\SIDM$ for the cooling time to be of order $H_0^{-1}$. At the boundary of the naive SIDM constraint, $\SIDM < 1 \text{cm}^2/\text{g}$, halos corresponding to $3-\sigma$ fluctuations could have a small enough cooling time to have undergone gravothermal collapse entirely, and assuming that the maximum stable mass of ADM stars is less than that of the protohalo mass, a black hole could form with mass on the order of the protohalo mass.  We will show in work to appear that the maximum stable halo mass is ${M_\text{max} \over M_\odot} \sim \left({\text{GeV} \over \bar m_X}\right)^2$ and so we expect black holes to form only if $\bar m_X \gtrsim \text{MeV}$.   On the other hand, we expect black holes seeded by DM fusion cooling to be vanishingly rare if $\SIDM < 0.1\, \cm^2/\gram$. 
There may be additional constraints or signatures from indirect detection or gravitational waves, which we reserve for future work.  

\section{Large Nugget Synthesis through a Bottleneck} 
\label{sec:bottleneck}

Without a strong bottleneck at low nugget size, synthesis proceeds through fusion of pairs of similarly sized nuggets until the reaction rate freezes out due to depleted number density---at typical nugget size $\kfo$.  In this scenario, we have seen above that achieving $\kfo \gtrsim 10^{17}$ requires fine-tuning in regions of parameter space not clearly ruled out by SIDM constraints. As discussed in \cite{Hardy:2014mqa,Gresham:2017cvl}, it is actually possible to synthesize larger nuggets if they are built up through capture of a dominant population of much smaller nuggets that persists because of a strong bottleneck at low $N$. Such a bottleneck could occur if, {\em e.g.}, both $\DN[3]$ and $\DN[4]$ were unstable. But a very small fraction of DM could squeeze through the bottleneck due to, e.g., a $3$-body interaction producing $\DN[6]$. Then the few nuggets that squeezed through the bottleneck could grow by capturing small nuggets. If squeezing through the bottleneck is sufficiently rare, the small nugget density controlling the capture rate can remain essentially constant even as the size of large nuggets increases many fold. This allows for freeze-out of nugget capture to occur at larger $N$.

In particular, we showed in Ref.~\cite{Gresham:2017zqi} (see also \cite{Hardy:2014mqa}) that the characteristic nugget size exiting synthesis in this bottleneck scenario is given by,
\begin{align}
\kfobn \approx \gamma_*^{3} \qquad \text{with} \qquad \gamma_* \sim \left[ {n_X  \over H} \; \pi \left({4 \pi n_\text{sat} \over 3}\right)^{-2/3}  \langle v {\cal T} \rangle \right]_{t_\text{syn}}  \label{eq: nfo bn}
\end{align}
where ${\cal T}$ is a velocity-dependent transmission factor caused by a possibly abrupt change of the effective mass of the constituent inside a saturated nugget.  For interactions between two saturated nuggets, ${\cal T}=1$ since the effective constituent masses are roughly the same. When ${\cal T}=1$, $\gamma_*$ and $\gamma$ in \Eq{eq:synsize} are the same, and we see that $\kfo^* \sim (\kfo)^{5/2}$; potentially \emph{much} bigger nuggets can be synthesized in the bottleneck scenario. 
However, strong bottlenecks tend to occur at small $N$, which is typically much smaller than the saturation size. Thus, the effective masses between the small and large nuggets are significantly different, leading to a transmission factor $ \langle v {\cal T} \rangle \sim v^2$ that can suppress the fusion rate (see \Eq{eq:fus}), making the contrast in size slightly less stark. In this scenario, we note that,
\beq
\gamma_* \sim 10^{10}  \left({g_*(T_\text{syn}) \over 10}\right)^{1/2} \bigg( \frac{1 \; \GeV}{\mxbar}\bigg)^{2}\bigg( \frac{ \mxbar^3}{n_{\sat}}\bigg)^{2/3} \bigg(\frac{T_{\rm syn}}{\mxbar} \bigg)^{2} \bigg(\frac{\mxbar}{M_\text{BN}} \bigg)
\eeq
where $M_\text{BN}$ is the mass of the dominant DM species (at the bottleneck) and we have taken $\langle {\cal T} v \rangle \sim v^2 \sim {T_\text{syn} \over M_\text{BN}}$.

It is also important to note that the approximation \Eq{eq: nfo bn} breaks down when the fraction of total dark number density in large nuggets approaches 1. If $p$ is the probability of a given nugget to squeeze through the bottleneck at the beginning of synthesis, and $N$ is the (rare) large nugget size, then this breakdown occurs when $p N \sim {\cal O}(1)$. At this point, fusion could continue through pairs of large nuggets. All told, in the bottleneck scenario we find
\begin{align}
  \kfo^* \simeq \min \left\{ 
\gamma_*^3, \, \frac{1}{p} + \gamma^{\frac{6}{5}}
\right\}\,,
\label{eq:NfoBN}
\end{align}
where $\gamma$ is the remaining interaction time after all the small nuggets are depleted, and it is computed by using \Eq{eq:synsize} with $T_\syn$ replaced by the temperature where the transition to large-large nugget fusion occurs.  One can easily check that nugget freeze-out size saturates to $1/p$ when all the small nuggets in the Hubble volume are captured onto the large nugget nucleation sites.  A lower bound on $p$ is obtained by requiring at least one nucleation site in a Hubble volume, which corresponds to the requirement $p > H(T_{\rm syn})^3/n_{^2X}(T_{\rm syn})$.  This condition will be easily satisfied over the entire parameter space we are interested in.

\begin{figure}
\includegraphics[width=0.45\textwidth]{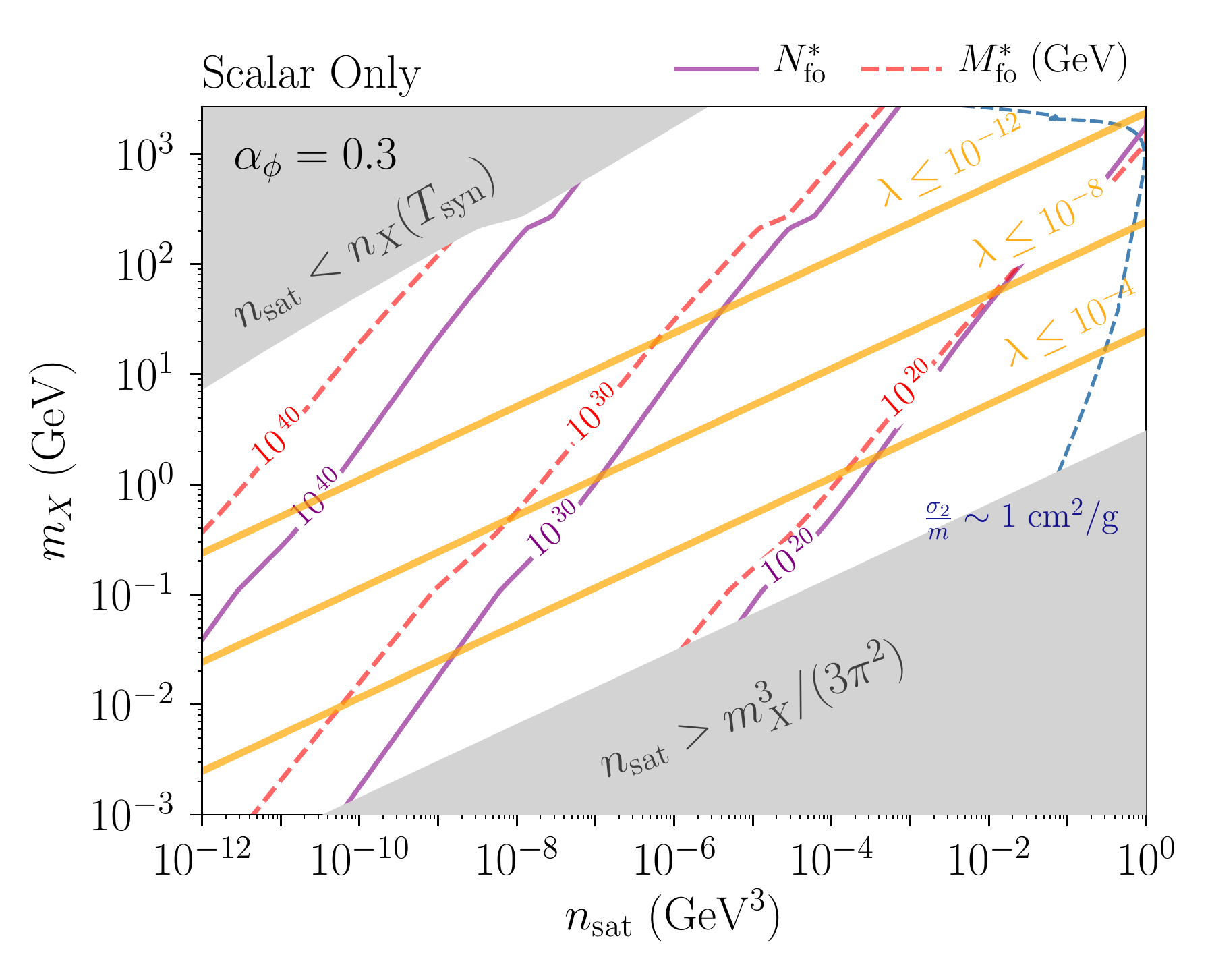} ~~\includegraphics[width=0.45\textwidth]{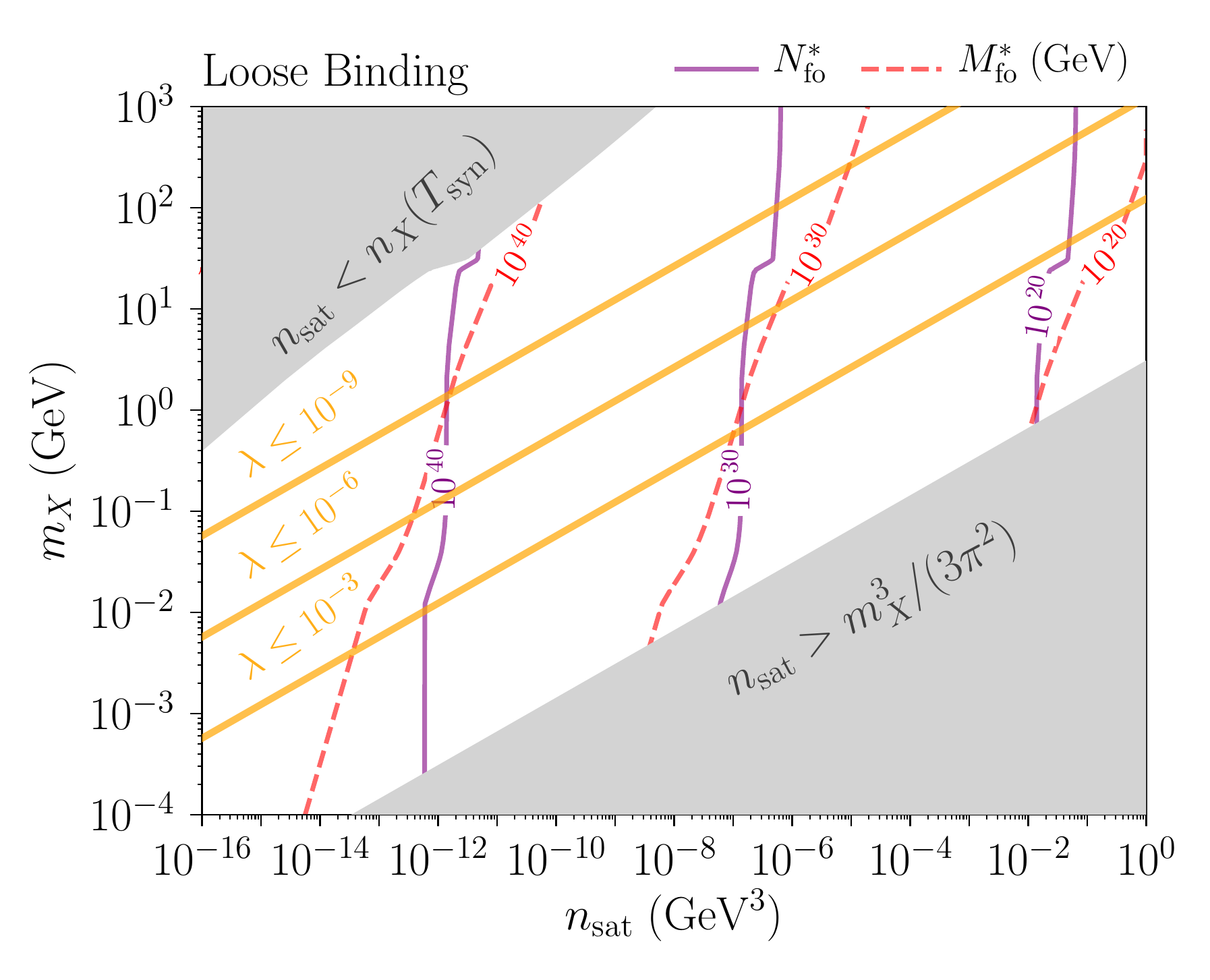}
\caption{Nugget size constraints for a scalar-binding model [left] and a benchmark loose-binding model ($\mxbar = 0.9~m_X$) [right], when a bottleneck is present. The dashed red (solid purple) contours indicate the characteristic mass (size) of the nuggets exiting synthesis, where $T_{\rm syn} = {\BE}_2/30 = \alpha_\phi^2 m_X/120$ in the left panel and $T_{\rm syn} = m_X/150$ in the right panel. \emph{Left:} The region left of the dashed blue line is excluded by the SIDM constraint $\SIDM < 1 \; \cm^2/g$ assuming that the dominant form of DM (by mass) is $\DN[2]$. 
\emph{Right:} A similar SIDM curve is not shown as it is strongly model dependent. On both plots, the solid orange curves indicate contours of constant scalar quartic. Small $\lambda$ values may imply fine-tuning. [See Eqs.~\eqref{eq: CV Cphi definitions}, \eqref{eq:lambda_max} and surrounding discussion.] It is possible to evade all SIDM constraints if synthesis ends when all small nuggets are depleted, in which case  $\kfo^{*}\sim p^{-1}$ as described in \Eq{eq:NfoBN}. For scenarios where $p$ (the probability to pass the bottleneck) is large enough such that small nuggets are quickly depleted and fusion is quickly dominated by large nuggets fusion, the standard analysis in \Sec{subsec:synthesis} applies, and we refer to \Fig{fig:no_bottleneck} for the relevant parameter space.}
\label{fig:bottleneck}
\end{figure}

Compared to the case without a bottleneck, the SIDM constraints with a bottleneck are much more model dependent. For instance, if small nuggets remain the dominant DM component in the late Universe, increased number density along with the fact that small nugget scattering may be effectively long-range can severely limit the parameter space. If all the small nuggets are fused into large ones, the SIDM constraints scale as ${\kfo^*}^{-\frac{1}{3}}$ which will depend on $p$ as in \Eq{eq:NfoBN}.  The left panel of \Fig{fig:bottleneck} shows an example of the relevant constraints in the scalar only model.  We took $T_\text{syn} = \BE_2/30 =  \alpha_\phi^2 m_X / 120$ for computing $\kfo^*$ and $M^*_{\rm fo}$, as in \Fig{fig:no_bottleneck}.  The blue dashed curve is the SIDM bound assuming $\DN[2]$ remains the dominant DM component. To model $\DN[2]$-$\DN[2]$ scattering interactions we assumed $\DN[2]$ is a point-like particle and have used the transfer cross section for an attractive potential (in the classical regime) as given in \cite{Tulin:2012wi,Tulin:2013teo} with $\alpha_X = 4 \alpha_\phi$. The SIDM constraint here rules out a majority of the parameter space and limits $\kfo^* \lesssim 10^{10}$ and $M_{\rm fo}^*\lesssim 10^{10}$ GeV.
 The right panel of \Fig{fig:bottleneck} shows a similar parameter space for a benchmark loose binding model, where $\mxbar = 0.9~m_X$, and we have again taken $T_{\rm syn}=m_X /150$ as in \Fig{fig:no_bottleneck}.  The SIDM constraint is model-dependent in this case and thus not shown in the figure. In the scenario where synthesis ends once small nuggets are depleted, $\kfo^*$ is maximized to be $p^{-1}$; in this case $\SIDM \lesssim 10^{-3} \; \cm^3/\gram$ is always satisfied in the available parameter space in \Fig{fig:bottleneck}.

Despite the lack of general SIDM constraint, there are self-consistency constraints that can become important when $m_X$ or $n_\sat$ is small.  
In particular, our estimates for synthesized size assume that aggregation proceeds primarily through two-body interactions.
This approximation will break down if $n_\text{sat}$ is comparable to or smaller than $n_X$ at any point during synthesis; if $n_\text{sat}\lesssim n_X(T_\text{syn})$, the Universe will begin as one single nugget and a phase transition will occur at some point, causing fragmentation into nuggets with sizes of order the Hubble size.
Such a synthesis mechanism could be interesting but is beyond the scope of this work. Requiring $n_\sat \gtrsim n_X(T_\syn)$ gives
 \begin{align}
\frac{n_\sat}{\mxbar^3}
\gtrsim 10^{-9} \;
\bigg( \frac{1\; \GeV}{\mxbar}\bigg)
\bigg( \frac{T_\syn}{\mxbar}\bigg)^{3} \,.
\end{align} 
We see that such an inequality is generally satisfied unless $\mxbar$ is very small.

\section{Conclusions}

We have studied the cosmology of ADM nuggets, and found several unique and generic signatures.  First, ADM nugget interactions are highly inelastic and exothermic---nuggets behave like clay putty when interacting, forming a compound state which then decays to the ground state through mediator or small nugget emission.  This means that most of the kinetic energy is lost in an interaction, implying a very effective cooling process in the late Universe.  This, combined with their {\em huge} size, gives rise to very efficient processes for changing the shape of DM halos, in particular in the core of a galaxy.  Such DM can efficiently feed the black hole in the galactic center.  On the flip side, requiring that our halo not be too greatly affected places an effective upper limit on the nugget size of around $10^{16} \mbox{ GeV}$, for models that are not too fine-tuned ($\lambda \gtrsim 10^{-3}$), as shown in \Fig{fig:Nmax_Mmax}.  The highly inelastic and exothermic nature of ADM nugget interactions means that many force mediators are emitted in the process of the compound state relaxing to the ground state; if the radiated force mediators decay to the SM (via, {\em e.g.}, mixing with the Higgs), we found that the parameter space in severely constrained, and the constituent masses must be quite heavy.  Lastly, the fusion processes allow for the formation of dark stars, though we find that once self-interaction and galactic core constraints are satisfied, these stars are formed only very rarely.

Large dark nuggets will form in the absence of a long range repulsive force (such as provided by electromagnetism in the standard model) given a sufficiently large attractive self-coupling (typically when $\alpha_\phi  > \alpha_V {m_\phi^2 /m_X^2}$ and $\alpha_\phi  > 50\, {m_\phi^2 /m_X^2}$). The generic presence of the bound states in models of ADM, as well as their qualitatively different astrophysical and experimental signatures from those of elementary particle DM, makes them ripe for further study. 

\section*{Acknowledgments}

We thank Guido D'Amico, Paolo Panci, Haibo Yu, and Yiming Zhong for very helpful discussions on gravothermal collapse in the presence of inelastic interactions, Dorota Grabowska, Ahmet Coskuner and Simon Knapen for collaboration on prospects to detect nuggets in direct detection experiments, and Ethan Neil and Matthew McCullough for discussions on low-$N$ bottlenecks.  HKL and KZ are supported by the DoE under contract No. DE-AC02-05CH11231. MG is supported by the NSF under award No.~1719780. 

\appendix

\section{Why Nugget Synthesis is Different from Standard Model Synthesis}
\label{app:sm_bottleneck}

Large nuclei are not synthesized in the SM.  Here we argue that {\em much} larger nuclei would be synthesized in the absence of the Coulomb force.  We further argue that in the absence of the Coulomb force, a small change in the structure of the dark sector could imply the absence of a bottleneck.

\subsection{In the Absence of a Bottleneck}

We first consider synthesis in the absence of a bottleneck, but with the presence of the Coulomb force.
With no bottleneck, the size, $N$, of a typical bound state evolves as
\beq
{\Delta N \over \Delta t} \sim {N \over (n \sigma v)^{-1}} \rightarrow {d N \over dt} = N \sigma_N {n_N} v_N = N \sigma_0 N^{2/3} e^{- \alpha N^2 / v_N} {n_X \over N} v_N
\eeq where the exponential term in the cross section characterizes the Coulomb barrier. We will assume that $v_N$ scales as $v_N = v_\circ N^{-1/2}$.
It is convenient to define the dimensionless time scale as in Eq.~(\ref{eq:synsize})
\beq
{d \gamma \over d t} = \sigma_0 n_X v_\circ. 
\eeq
Then the evolution equations for average size $N$ are
\beq
{d N \over d\gamma} = N^{1/6} e^{-{\alpha \over v_\circ(\gamma)} N^{5/2}}.
\eeq
 If $v_\circ'(\gamma_\circ) \Delta \gamma$ is very small compared to $v_\circ(\gamma_\circ)$, then, defining $\beta = \alpha / v_\circ(\gamma_\circ)$ we have 
\beq
\gamma = \int N^{-1/6} e^{\beta N^{5/2}} dN 
\approx \bigg\{ 
\begin{array}{l l} 
{2 \over 5}\beta^{-1} N^{-5/3} e^{\beta N^{5/2}} & \text{if}~\beta N^{5/2} \gtrsim 2 \\
{6 \over 5} N^{5/6} & \text{if}~\beta N^{5/2} \ll 1
\end{array}.
\eeq 
In the SM, with the synthesis starting around 0.1 MeV due to the deuterium bottleneck, $\gamma \approx 3000$, $v_\circ \sim { \sqrt{T_\text{BBN}/\text{GeV}}} \sim 10^{-2}$.  Solving for $N$, one obtains $N \approx 2.56$ due to the strong exponential dependence, indicating the inefficiency of SM synthesis (and correctly predicting that synthesis stops at around $Z=2$, helium).  On the other hand, if the Coulomb barrier were absent, the same calculation would predict $N \sim 10^4$.  

\subsection{In the Presence of a Bottleneck}

If there is a bottleneck at low $N$, large nuggets can build up by capture of small bound states on sparse nucleation sites that squeeze through the bottleneck. Suppose the bottleneck is at size $k$. The the size of the nucleation sites grows as
\beq
{d N \over d t} = k n_k \sigma_{k N} v_k.
\eeq
Taking $k n_k = n_X(1 - p N)$ and   $\sigma_{k N} = \sigma_\circ N^{2/3} e^{- \alpha N / v_k} f(v_k)$ with $f(v_k)$ a possible suppression factor due to quantum reflection effects we have
\beq
{d N \over d \gamma^*} = (1- p N)N^{2/3} e^{- \alpha N / v_k}
\eeq where 
\beq
{d \gamma^* \over d t} \equiv \sigma_\circ n_X v_k f(v_k).
\eeq
Here $p$ is the probability of squeezing through the bottleneck.
In the limit where $v_k$ is approximately constant over the interaction timescale and if $p N \ll 1$ then 
\beq
{d N \over d \gamma_*} = N^{2/3} e^{-\beta N}
\eeq so that 
\beq
\gamma^* = \int N^{-2/3} e^{\beta N} dN \approx 
\bigg\{ 
\begin{array}{l l} 
\beta^{-1} N^{-2/3} e^{\beta N} & \text{if}~\beta N \gtrsim 2 \\
3 N^{1/3} & \text{if}~\beta N \ll 1
\end{array}
\eeq
Again using SM as an example, with $\gamma^*\sim 3000$ and $\beta = 1$ one predicts $N \sim 9.5$. This estimate confirms that a sparse population of $A>4$ nuclei could not grow substantially through capture of helium during BBN. If the Coulomb barrier were absent, however, one would predict $N \sim 10^9$. 

\subsection{Are Bottlenecks Present in the Dark Sector?}

\def\he{{}^4\text{He}}
\def\hethree{{}^3\text{He}}
\def\hefive{{}^5\text{He}}

The synthesis of large $N$ nuggets, and their abundance in the late Universe, depends crucially on the presence or absence of a bottleneck at small $N$.  From analogue with the SM, one might think that bottlenecks are a generic feature of bound states.  Here we argue that in the absence of the Coulomb barrier, and with very slight tweaks to the strong interaction physics that determines the $^4$He and $^8$Be binding energies, bottlenecks would be absent in the SM.  

Our estimates above show that the Coulomb barrier is primarily responsible for the BBN bottleneck at ${}^4\text{He}$. But if the deuterium bottleneck were not so strong so that BBN happened slightly earlier, this barrier would not be so huge. At higher temperatures, SM fusion is suppressed also because ${}^8\text{Be}$ is unstable. 
This is tied to the fact that as a  ``doubly magic'' nucleus with both protons and both neutrons paired, filling the $L=0$ orbitals, $\he$ is especially tightly bound.  In contrast, consider the unstable ${}^5\text{He}$ and ${}^8\text{Be}$ nuclei:
\bit
\item The third neutron in $\hefive$ is unpaired and in the $L=1$ orbital (2nd ``shell''), leading to a total $\hefive$ binding energy less than that of $\he$. Thus $\hefive$ rapidly decays to $\he + n$. 
\item All neutrons and protons in ${}^8\text{Be}$ are paired, though the last two protons (neutrons) fill only 1/3 of the $1p$ (2nd shell) states. The ${}^8\text{Be}$ binding energy per particle is the smallest of any isotope with $A=5$ to $11$, but it is \emph{just barely} smaller than that of $\he$.\footnote{${\BE(\he) / 4 \over \BE({}^8\text{Be}) / 8} - 1 = 0.0016 $.} This means ${}^8\text{Be}$ decays rather quickly (through the strong interaction) to $ \he + \he$.
\eit

Now consider the existence of analogous bottlenecks in nugget synthesis. First, there will be no obstruction due to a Coulomb barrier. One could expect, however, for the angular momentum-dependent pairing and shell filling effects to modify the behavior of binding energy per particle especially at low $N$, which in principle could destabilize nuggets at certain $N$. Specifically, we might expect the binding energy per particle to  shift down for odd-$N$ nuggets (with an unpaired constituent) relative to even-$N$ nuggets (with all constituents paired) or to have upward fluctuations in binding energy per particle at the magic numbers ($N\approx2, 8, 20, \ldots $).\footnote{One can expect the larger-$N$ magic numbers to be different than the magic numbers for nuclei because the strength of the  spin-orbit interaction, which leads to reordering of shell energies, will generically be different.}  The analog to an absence of stable $A=5$ and $A=8$ states for nuclei would be an absence of stable $N=3$ and $N=4$ nuggets; this would require the total binding energy of $\DN[2]$ to be larger than that of $\DN[3]$, and the binding energy per particle for $\DN[4]$ to be (even slightly) smaller than that for $\DN[2]$. Unstable $\DN[3]$ and $\DN[4]$ would constitute a strong bottleneck to fusion of larger nuggets; the majority of DM could exist as $\DN[2]$ after early-Universe synthesis. However, if only one of $\DN[3]$ or $\DN[4]$ were unstable, we could expect fusion to proceed to large $N$. 

To definitively answer the question of the small-$N$ structure of bottlenecks requires detailed numerical calculations (see e.g.~\cite{Kim:2009sr,Pahlavani:2010zzb,Beane:2012vq,Barnea:2013uqa,Detmold:2014kba,Orginos:2015aya}), 
though given how close the $A=5$ and $A=8$ nuclei in the SM totter towards stability, it is not hard to imagine that a dark sector with a different structure could provide for the absence of low-$N$ bottlenecks.

\section{Saturation Properties from Relativistic Mean Field Theory}
\label{app:saturation}

With only the scalar and vector contributing to large-$N$ nugget properties, our EFT mimics the same behavior as the $\sigma$-$\omega$ model of nuclear physics. In \cite{Gresham:2017zqi} we examined the saturation properties of nuggets given only a scalar mediator and a quartic scalar potential as well as for scalar and vector mediators but no mediator potential. Here we sketch the derivation of saturation properties of nuggets for completeness. We omit many details that can be found in textbooks such as \cite{Walecka:1995mi, Glendenning:1997wn}.

In mean field calculations the mediator fields are set to their expectation values and treated classically.  It is useful to rewrite the Lagrangian using an alternative parameterization of the couplings and dimensionless fields,
\begin{align}
  C_{\phi, V}^2 \equiv \frac{2 g_{\rm dof}\alpha_{\phi, V}}{3\pi}\frac{m_X^2}{m_{\phi,V}^2}
\qquad \sphi \equiv \frac{g_\phi \langle \phi \rangle}{m_X} \qquad v_\mu \equiv \frac{g_V \langle V_\mu \rangle}{m_X}\,.
\end{align}
The potential term can be rewritten as
\begin{align}
  V(\sphi) \equiv \frac{g_{\rm dof} m_X^4}{6\pi^2}W(\sphi).
\end{align}
Here $g_\text{dof}$ is the number of degrees of freedom of the fermionic constituent. In a model with flavor symmetry, $g_\text{dof} = 2 n_\text{flavors}$. We take $g_\text{dof}=2$ in all numerical calculations but include $g_\text{dof}$ explicitly here partly to compare to the nuclear $\sigma$-$\omega$ model in which $g_\text{dof} = 4$.

The equations of motion for the vector and scalar fields in the saturation limit become 
\begin{align}
\sV_\mu &= \delta_{0,\mu} \frac{C_V^2 k_F^3}{m_X^3}
\label{eq:eom_v}
\\
 \sphi &= - C_\phi^2W'(\sphi) + 3 C_\phi^2 \int_0^{k_F/m_X}dx \frac{x^2 (1- \sphi)}{\sqrt{x^2+(1-\sphi)^2}} \,, 
\label{eq:eom_phi}
\end{align} and the equilibrium (zero pressure) condition is
\begin{align}
 p\left(\frac{g_{\rm dof} m_X^4}{6\pi^2} \right)^{-1} &= 
-\frac{\sphi^2}{2C_\phi^2} + \frac{\sV_0}{2}\left(\frac{k_F}{m_X} \right)^3 - W(\sphi) + \int_0^{k_F/m_X}dx \frac{x^4}{\sqrt{x^2 + (1-\sphi)^2}}=0\,.
\label{eq:eom_p}  
\end{align}
Here $k_F$ is the Fermi momentum and $m_* = m_X (1 - \sphi)$ is the effective mass of the fermion constituents. Solving Eq.~\ref{eq:eom_v}-\ref{eq:eom_p}, one can obtain the mean field values $(\sV_0, \sphi, k_F)$. The physical properties can then be derived 
\begin{align}
  n_{\sat} = \frac{g_{\rm dof}k_F^3}{6\pi^2} \qquad
  \frac{\mxbar}{m_X}= \sV_0 + \sqrt{ (1-\sphi)^2 + \left(\frac{k_F}{m_X} \right)^2}  \,,
\end{align}
where $n_{\sat}$ is the nugget number density and $\mxbar$ is the energy per nugget number. One immediately sees
\beq
{n_\text{sat} \over \bar m_X^3} \leq {g_\text{dof} \over 2}{1 \over 3 \pi^2}{k_F^3 \over \left(k_F^2 + m_*^2 \right)^{3/2}} \leq {g_\text{dof} \over 2}{1 \over 3 \pi^2}. \label{eq: properties}
\eeq
The bound on $n_\text{sat} / \bar m_X^3$ is saturated when $C_V^2=0$ and in the ultrarelativistic limit, where $m_*/k_F \rightarrow 0$. Generically, the presence of a vector field increases the pressure, and lowers both the nugget binding energy and saturation density. 

Defining $W_\text{eff}(\sphi)  \equiv  \frac{\sphi^2}{2C_\phi^2} + W(\sphi) $, and $z  = k_F/m_X$ and substituting in for $v_\mu$, the equations for saturation become
\begin{align}
 W_\text{eff}'(\sphi) &=3 \int_0^{z}dx \frac{x^2 (1- \sphi)}{\sqrt{x^2+(1-\sphi)^2}} \,, 
\label{eq:eom_phi}
\\
W_\text{eff}(\sphi) &= 
 \frac{C_V^2}{2} z^6 +  \int_0^{z}dx \frac{x^4}{\sqrt{x^2 + (1-\sphi)^2}}\,.
\label{eq:eom_p}
\end{align}
 For binding to occur, one must have $\bar m_X < m_X$, so by \Eq{eq: properties}, binding requires $z < 1$ and  $C_V^2 z^3 < 1$. Also note that $0 \leq 1 - \sphi < 1$.

 \Eq{eq:eom_phi}-(\ref{eq:eom_p}) both go to zero as $z\rightarrow 0$. Thus, small saturation densities ($z \ll 1$) requires small $W_\text{eff}(\sphi)$ and $W_\text{eff}'(\sphi)$. This can be achieved either by making the coefficients of $W_\text{eff}(\sphi)$ very small, or requiring $\sphi  \ll 1$. However, as $\bar m_X / m_X > |1-\sphi|$, it is difficult to achieve binding in this limit. In the ultrarelativistic limit $(z \gg 1)$, the equations simplify, and we will show that a consistent limit can be achieved as long as $W_\text{eff}(1) \ll 1$.

\paragraph*{Relativistic limit.}

Suppose saturation occurs in the ultrarelativistic limit, where $1 - \sphi \ll z$. We'll take the limit and then see when it is consistent. When $1 - \sphi \ll z$ we have
\begin{align}
W_\text{eff}'(\sphi)/(1-\sphi) &\approx {3 \over 2} z^2  \label{eq: saturation a} \\
W_\text{eff}(\sphi) &\approx {1 \over 2}C_V^2 z^6 + {1 \over 4} z^4. \label{eq: saturation b}
\end{align}
First of all, since, for binding to occur, we need $z < 1$, we can see that $1 - \sphi \ll 1$ in the ultrarelativistic limit, implying that $\sphi \approx 1$ and so $W_\text{eff}(\sphi) \approx W_\text{eff}(1) + W_\text{eff}'(1) \, (\sphi - 1) + \ldots$. Then Eqs.~\ref{eq:eom_phi} and \ref{eq:eom_p}  become
\begin{align}
W_\text{eff}'(1) &= \left( {3 \over 2} z^2 + W_\text{eff}''(1)  \right) (1 - \sphi) + W_\text{eff}'''(1) \, (1-\sphi)^2 + {\cal O}\left( (1 - \sphi)^2 z, (1-\sphi)^3 \right)   \label{eq: saturation a1} \\
W_\text{eff}(1) &= z^4 \left({1 \over 2}C_V^2 z^2 + {1 \over 4}\right) + W_\text{eff}'(1)\, (1 - \sphi) + {\cal O}(z^3 (1-\sphi), (1-\sphi)^2) . \label{eq: saturation b1}
\end{align}
For binding to occur,  $C_V^2 z^2 < {1 \over z}$ and $z < 1$. Therefore ${\cal O}(z^4) \leq W_\text{eff}(1) \leq {\cal O} (z^3) < 1$.   A consistency condition on $W_\text{eff}$ is
\beq
 {{2 \over 3} W_\text{eff}'(1) \over \left( 4 W_\text{eff}(1) \right)^{3/4}} \approx {1 - \sphi \over z \left( 1 + 2 C_V^2 z^2 \right)^{3/4}} < {1 - \sphi \over z } \ll 1.
\eeq which, noting that $W_\text{eff}$ and its derivatives evaluated at $1$ must be of the same order assuming positive coefficients, implies that $W_\text{eff}(1) \ll 1$ and therefore $z \ll 1$ is necessary for the limit to be consistent.  Therefore, we have 
\begin{align}
W_\text{eff}'(1) &\approx \left( {3 \over 2} z^2 \right) (1 - \sphi)   \label{eq: saturation a2} \\
W_\text{eff}(1) &\approx z^4 \left({1 \over 2}C_V^2 z^2 + {1 \over 4}\right)  \label{eq: saturation b2}
\end{align}
along with 
\beq
n_\text{sat} = {z^3 \over 3 \pi^2} \qquad {\bar m_X \over m_X} \approx z \left( C_V^2 z^2 + 1\right)
\eeq
where $(1 - \sphi) \ll z \ll 1$ and $C_V^2 z^2 < 1/z$. We see that saturation densities are small in this limit and a large range of binding energies is self-consistently achievable. Namely, $C_V^2 z^2 \ll 1$ corresponds to strongly bound nuggets and $C_V^2 z^3 \approx 1$ corresponds to weakly bound nuggets. 

\Eq{eq: saturation b2} is a cubic equation for $z^2$ whose solution is 
\beq
z^2 = {1 \over 6 C_V^2} \left[ \left( \sqrt{\xi^2 - 1} - \xi \right)^{1/3} + \left( \sqrt{\xi^2 - 1} - \xi \right)^{-1/3} - 1 \right] ~~ ; ~~ \xi \equiv 1 - 216 C_V^4 W_\text{eff}(1).  \label{eq: cubic solution}
\eeq
The following formulas describe the solution \Eq{eq: cubic solution} to within $33\%$ within the entire range:
\begin{align}
z^2 & = 2 \sqrt{ W_\text{eff}(1)}   &  \qquad  C_V^4 W_\text{eff}(1) & \lesssim 1/16 \\
z^2 & = \left( 2 W_\text{eff}(1) \over C_V^2 \right)^{1/3}  &  \qquad    C_V^4 W_\text{eff}(1) & \gtrsim 1/16 
\end{align}
and, correspondingly,
\beq
3 \pi^2 {n_\text{sat} \over m_X^3}  =  \Bigg\{ 
\begin{array}{l l}
{\left(2 \sqrt{W_\text{eff}(1)} \right)^{3/2} }  &  \qquad  C_V^4 W_\text{eff}(1) \lesssim 1/16 \\
 {\sqrt{2 W_\text{eff}(1) } /  C_V} & \qquad C_V^4 W_\text{eff}(1)  \gtrsim 1/16
\end{array} 
\eeq
and
\beq
{\bar m_X \over m_X} = \Bigg\{ 
\begin{array}{l l}
\left(2 \sqrt{W_\text{eff}(1)} \right)^{1/2} &  \qquad  C_V^4 W_\text{eff}(1) \lesssim 1/16 \\
 \left( {\sqrt{2 W_\text{eff}(1) }/ C_V}\right)^{1/3} \left( 1+(2 C_V^4 W_\text{eff}(1))^{1/3}\right) & \qquad C_V^4 W_\text{eff}(1)  \gtrsim 1/16
\end{array} \\
\eeq
with the consistency conditions,
\begin{align}
{ {2 \over 3} W_\text{eff}'(1) \over \left(2 \sqrt{W_\text{eff}(1)} \right)^{3/2}} &\ll 1 &  \qquad  [ C_V^4 W_\text{eff}(1) \lesssim 1/16] \\
C_V{ {2 \over 3} W_\text{eff}'(1) \over \left(\sqrt{2 W_\text{eff}(1)} \right)} &\ll 1  &  \qquad    [C_V^4 W_\text{eff}(1)  \gtrsim 1/16 ].
\end{align}

Very large $C_V$ can destabilize nuggets, corresponding to $\bar m_X / m_X = 1$. In the limit $ (2 C_V^4 W_\text{eff}(1))^{1/3} \gg 1$ we have ${\bar m_X \over m_X} \rightarrow C_V \sqrt{2 W_\text{eff}(1)}$ and thus we see the limit for binding
\beq
C_V^2 W_\text{eff}(1) < 1/2  \qquad \text{when} \qquad C_V^4 W_\text{eff}(1) \gg 1 \qquad  (\text{binding limit}).
\eeq

Let us redefine $C_{\phi}^{-2} \equiv 2 W_\text{eff}(1)$. We may invert the formulae for $n_\text{sat}$ and $\bar m_X$ to give $C_{\phi}^{-2} \equiv 2 W_\text{eff}(1)$ and $C_V^2$. We find,
\begin{multline}
C_{\phi}^{-2} = (r_0 m_X)^{-3} \left( {\bar m_X - r_0^{-1} \over m_X}  \right) \qquad \text{and} \qquad C_V^2 =(r_0 m_X)^{3} \left( {\bar m_X - r_0^{-1} \over m_X}  \right)  \\ \text{with}~~r_0^{-1}<\bar m_X < m_X \qquad \text{when} ~~ ( C_V^4 C_{\phi}^{-2})^{1/3} \sim   {\bar m_X r_0 - 1}   \gtrsim 1/2 \label{eq: cphi 1}
\end{multline} and 
\beq
C_{\phi}^{-2} \rightarrow {1 \over 2} \left({\bar m_X \over m_X}\right)^{4} = {1 \over 2} \left({ r_0^{-1} \over m_X}\right)^{4} \qquad \text{and} \qquad C_V^2 \rightarrow 0  \qquad \text{as}~~ r_0^{-1} \rightarrow \bar m_X \label{eq: cphi 2}
\eeq
where we have defined 
\beq
 {r_0^{-1} \equiv (3 \pi^2 n_\text{sat})^{1/3} } .
\eeq

Combining Eqs.~\ref{eq: cphi 1} and \ref{eq: cphi 2} we find 
\beq
C_{\phi}^{-2} \leq  3 \pi^2 {n_\text{sat} \over \bar m_X^3}  \left( \bar m_X \over m_X \right)^4  \left[1 - {1 \over 2}\left( {3 \pi^3 n_\text{sat} \over \bar m_X^3}\right)^{1/3} \right]. \
\eeq

\bibliography{ADMNuggetsLateUniverse,ADMNuggets,TL}

\end{document}